\documentclass[useAMS,usenatbib]{mn2e}
\usepackage{times}
\usepackage{graphicx,epsfig}

\title[Ram pressure stripping in Hickson groups] {Galaxy evolution in
  Hickson compact groups: The role of ram pressure stripping and
  strangulation} \author[J. Rasmussen et al.]{Jesper
  Rasmussen,$^{1}$\thanks{E-mail: jr@ociw.edu}\thanks{Chandra Fellow}
  Trevor J. Ponman,$^{2}$ Lourdes Verdes-Montenegro,$^{3}$ Min
  S. Yun$^{4}$ \newauthor and Sanchayeeta Borthakur$^{4}$\\
  $^{1}$Observatories of the Carnegie Institution, 813 Santa Barbara
  Street, Pasadena, CA 91101, USA\\ $^{2}$School of Physics and
  Astronomy, University of Birmingham, Edgbaston, Birmingham B15 2TT\\
  $^{3}$Instituto de Astrof\'{\i}sica de Andaluc\'{\i}a, CSIC, Apdo.\
  Correos 3004, E-18080 Granada, Spain\\ $^{4}$Astronomy Department,
  University of Massachusetts, Amherst, MA 01003, USA}

\begin{document} 
 
\date{} 
 
\pagerange{\pageref{firstpage}--\pageref{lastpage}} \pubyear{2008} 
 
\maketitle 
 
\label{firstpage} 
 
\begin{abstract}
  Galaxies in compact groups tend to be deficient in neutral hydrogen
  compared to isolated galaxies of similar optical properties. In
  order to investigate the role played by a hot intragroup medium
  (IGM) for the removal and destruction of H{\sc i} in these systems,
  we have performed a {\em Chandra} and {\em XMM-Newton} study of
  eight of the most H{\sc i}~deficient Hickson compact groups. Diffuse
  X-ray emission associated with an IGM is detected in four of the
  groups, suggesting that galaxy--IGM interactions are not the
  dominant mechanism driving cold gas out of the group members. No
  clear evidence is seen for any of the members being currently
  stripped of any hot gas, nor for galaxies to show enhanced nuclear
  X-ray activity in the X-ray bright or most H{\sc i}~deficient
  groups. Combining the inferred IGM distributions with analytical
  models of representative disc galaxies orbiting within each group,
  we estimate the H{\sc i} mass loss due to ram pressure and viscous
  stripping. While these processes are generally insufficient to
  explain observed H{\sc i}~deficiencies, they could still be
  important for H{\sc i} removal in the X-ray bright groups,
  potentially removing more than half of the ISM in the X-ray bright
  HCG\,97. Ram pressure may also have facilitated strangulation
  through the removal of galactic coronal gas. In X-ray undetected
  groups, tidal interactions could be playing a prominent role, but it
  remains an open question whether they can fully account for the
  observed H{\sc i}~deficiencies.
\end{abstract}
 
\begin{keywords} 
  galaxies: evolution --- galaxies: interactions --- galaxies: ISM ---
  X-rays: galaxies --- X-rays: galaxies: clusters
\end{keywords}

\section{Introduction}

The origin of the galaxy morphology-density relation is still one of
the most important unsolved problems in astrophysics. Not only are
spiral galaxies less common within dense cluster environments, but
those which are present tend to be deficient in H{\sc i}, and this
deficiency itself correlates with projected local galaxy density
(e.g., \citealt{giov85}). The mechanisms responsible for the changes
in the morphology and gas content of galaxies are unclear, with gas
stripping, tidal shocks, and galaxy interactions and mergers all
contenders.

Traditionally, clusters of galaxies have represented the environment
of choice for attempts to unravel the nature of the relevant
processes. Recent results, however, strongly suggest that the origin
of the environmental modification of galaxies which underpin the
morphology-density relation lies, not in the cores of galaxy clusters,
but in smaller groups and cluster outskirts. For example,
spectroscopic studies of the effects of the cluster environment on
galaxies (e.g., \citealt{lewi02}) show that the suppression of star
formation takes place in cluster outskirts rather than in the core,
and is modulated by local galaxy density. Moreover, X-ray bright
groups show a morphology-density relation {\it stronger} than that of
clusters \citep{hels03}, a result adding to the accumulating evidence
that cluster galaxies have often been `pre-processed' in groups prior
to their assembly into larger systems (see, e.g.,
\citealt{cort06}). In addition, it is becoming clear that processes
once thought to be exclusive to the cluster environment, such as ram
pressure stripping \citep*{rasm06} and strangulation \citep{kawa08},
may play a role also in much smaller systems. In order to elucidate
the origin of the morphology-density relation, it is therefore
necessary to study the processes acting on galaxies within groups.

The compact groups in the catalogue of \citet{hick82} offer
particularly interesting opportunities in this regard. Many of these
groups are spiral-rich, but their galaxy population is, on average,
deficient in H{\sc i} by a factor of $\sim 2$ compared to loose groups
\citep{will87}. Furthermore, some of these groups have exceptionally
compact galaxy configurations, and so may represent pre-virialisation
systems close to maximum collapse, in which galaxies are suffering
strong environmental modification but have yet to be converted into
early-types. While recent work has identified tidal interactions and
mergers as playing an important role in the morphological
transformation of spirals in some compact groups \citep{cozi07}, it is
still unclear what is causing the observed H{\sc i}~deficiencies and
to what extent this is related to the processes modifying the stellar
component of the group members.

\citet{verd01} presented a detailed study of the H{\sc i} content of a
sample of 72 Hickson compact groups (HCGs), with integrated H{\sc i}
masses from single dish measurements, and detailed Very Large Array
(VLA) mapping of a subsample. Defining H{\sc i}~deficiency
$\Delta_{\rm HI}$ as
\begin{equation}
  \Delta_{\rm HI} \equiv {\rm log} M_{\rm{HI,pred}} - 
  {\rm log} M_{\rm{HI,obs}},
\label{eq,hi_def}
\end{equation}
where $M_{\rm{HI,obs}}$ is the observed H{\sc i} mass of the group
galaxies and $M_{\rm{HI,pred}}$ is that predicted for isolated
galaxies of similar morphology and optical luminosity \citep{hayn84},
this work confirmed the deficiency in HCGs, and allowed a search for
correlations between deficiency and other group properties. One of the
strongest relationships found was that with detectable intergalactic
X-ray emission; almost half of the groups with significant H{\sc i}
deficiency showed diffuse intragroup X-ray emission in the {\em
  ROSAT} survey of \citet[hereafter P96]{ponm96}.  More recently,
similar results have been reported also for groups outside Hickson's
catalogue \citep{seng06}.

The increased prevalence of X-ray detected systems among H{\sc i}
deficient (compact) groups may suggest a picture whereby H{\sc i} is
stripped from spiral galaxies within virialising groups, and then
destroyed due to heating by a surrounding hot intragroup medium
(IGM). However, for most of the compact groups in the \citet{verd01}
sample, {\em ROSAT} data were either not available or of insufficient
quality to permit any detailed study of the relationship between the
hot and cold gas. With only very shallow {\em ROSAT} All-Sky Survey
data at their disposal for a number of these systems, \citet{verd01}
could not establish the amount and properties of any hot IGM in these
groups.

In order to explore the processes destroying H{\sc i} in these
systems, we have therefore embarked on a programme to obtain high
quality X-ray and radio data for the most H{\sc i}~deficient compact
groups, adding in existing archival X-ray data wherever relevant.
Since ram pressure stripping can only operate where a significant IGM
is present, while tidal stripping of gas requires only that galaxies
be interacting, a key discriminator for the mechanism of H{\sc i}
removal from galaxies is whether or not there is a correlation between
H{\sc i}~deficiency and the properties of a hot IGM, and in particular
whether a hot IGM is present in the highly H{\sc i}~deficient
systems. Clarifying these issues represents the goal of the present
study, in an attempt to shed light on the role played by galaxy--IGM
interactions in destroying H{\sc i} and establishing the
morphology-density relation in these compact groups. The focus of this
paper thus rests mainly on the X-ray properties of any hot
intergalactic gas within the groups, while a forthcoming companion
paper (Verdes-Montenegro et al., in prep.) will discuss in more detail
the H{\sc i} and radio continuum emission in the groups, the detailed
relationship between X-ray and H{\sc i} morphology, and the
multi-wavelength properties of the individual group galaxies.

In Sections~\ref{sec,sample} and \ref{sec,data} we outline the sample
selection and X-ray data analysis, respectively.
Section~\ref{sec,results} presents the results for the X-ray
properties of the hot gas and galaxies within each group, and compares
the derived IGM properties to the observed H{\sc i}~deficiencies. In
Section~\ref{sec,model} we construct an analytical model of a
representative late-type galaxy orbiting within the derived
gravitational potential of each X-ray bright group, allowing us to
evaluate the importance of ram pressure and viscous stripping for
H{\sc i} removal. The results are discussed in
Section~\ref{sec,discus} and summarized along with the main
conclusions in Section~\ref{sec,conclus}.

A Hubble constant of 73~km~s$^{-1}$~Mpc$^{-1}$ is assumed throughout.
Unless otherwise stated, all errors are quoted at the 68~per~cent
confidence level.

\section{Group sample and observations}\label{sec,sample}

Our broad aim is to establish the relationship between the hot and
cold gas in H{\sc i}~deficient groups, and explore the processes of
gas removal operating within them. Our sample is therefore based on
that studied by \citet{verd01}, from which we selected all HCGs which
are highly deficient in H{\sc i} ($\Delta_{\rm HI} > 0.5$) based on
Very Large Array (VLA) H{\sc i} data, contain four or more group
galaxies, and lie at a distance $D<100$~Mpc. This yielded an initial
sample of 11 groups, which will be discussed in its entirety in
Verdes-Montenegro et~al.\ (in prep.). VLA H{\sc i} mapping has been
completed for all 11 systems, along with follow-up Green Bank
Telescope (GBT) observations for the groups discussed in this paper.
Compared to the VLA, the single-dish GBT is more sensitive to
extended, smoothly distributed emission that may otherwise be filtered
out by the VLA interferometer. Including the GBT data thus provides a
more complete census of the H{\sc i} content in the groups.  For
HCG\,48, included in the initial sample, the additional gas detected
by GBT indicates that this group is not H{\sc i}~deficient after
all. Consequently, this group was omitted from the sample for the
purpose of the present study.

Of the remaining systems, the eight included in this paper are those
for which {\em Chandra} or {\em XMM-Newton} X-ray data are currently
available.  Four of these eight groups are exceptionally compact, with
major galaxies (as catalogued by \citealt{hick82}) contained within a
circle of radius $\sim 2$~arcmin, and thus requiring {\em Chandra}
data to resolve their X-ray structure in the crucial region.  We note
that these groups are not generally mature, X-ray bright systems
dominated by early-type galaxies. Such groups may contain little H{\sc
i}, but the H{\sc i} content of their galaxies as predicted from the
\citet{hayn84} results mentioned above is also small (zero for
ellipticals and modest for lenticulars). Such groups are therefore not
usually H{\sc i} {\em deficient} according to our definition in
equation~(\ref{eq,hi_def}).  Our chosen systems have H{\sc i} content
far below that expected for their galaxy contents, and are therefore
those in which the processes which destroy H{\sc i} should be in
active, or very recent, operation.

The observation log for the X-ray data considered in this paper is
presented in Table~\ref{tab,X}, detailing the pointing coordinates for
each observation (for archival observations not necessarily identical
to the optical group centre), group distance $D$, 
the observing instrument, date, and mode, along
with cleaned exposure times $t_{\rm exp}$ for each camera, and the
Galactic absorbing column density $N_{\rm H}$ from \citet{dick90} as
adopted in the X-ray spectral analysis.

\begin{table*} 
 \centering
 \begin{minipage}{139mm} 
   \caption{Log of available X-ray observations. Group luminosity
    distances $D$ for the adopted value of $H_0$ were taken from the
    NASA/IPAC Extragalactic Database (NED). Column~7 specifies the
    frame mode (full frame/extended full frame) and optical blocking
    filter for {\em XMM}, and ACIS CCD aimpoint and telemetry mode
    (Faint/Very Faint) for {\em Chandra}.}
  \label{tab,X} 
  \begin{tabular}{@{}lccclcccc@{}} \hline 
   \multicolumn{1}{l}{Group} & 
\multicolumn{1}{c}{RA} &
\multicolumn{1}{c}{Dec} &
\multicolumn{1}{c}{$D$} &
\multicolumn{1}{c}{{\em Chandra}/} &
\multicolumn{1}{c}{Obs.\ date} &
\multicolumn{1}{c}{Obs.\ mode} &
\multicolumn{1}{c}{$t_{\rm exp}$} &
\multicolumn{1}{c}{$N_{\rm H}$} \\
   & (J2000) & (J2000) & (Mpc) & {\hspace{1.4mm} {\em XMM}} & (yyyy-mm-dd)&  
   &(ks) {\hspace{0mm}} & ($10^{20}$ cm$^{-2}$)\\ \hline
HCG\,7   & 00 39 13.5 & $+00$ 51 49.3 & 54 & {\em XMM} pn & 2004-12-26 
        & FF Thin & 25.9 & 2.24  \\
 \ldots & \ldots & \ldots & \ldots & {\em XMM} m1 &\ldots & FF Thin & 35.1 
        &\ldots \\
 \ldots & \ldots & \ldots & \ldots & {\em XMM} m2 &\ldots & FF Thin & 35.9
        & \ldots \\
HCG\,15  & 02 07 39.0 & $+02$ 08 18.0 & 92 & {\em XMM} pn & 2002-01-10 
        & EFF Thin & 23.3 & 3.20  \\
 \ldots & \ldots & \ldots & \ldots & {\em XMM} m1&\ldots  & FF Thin & 31.0 
        &\ldots \\
 \ldots & \ldots & \ldots & \ldots & {\em XMM} m2 &\ldots & FF Thin & 31.1 
        &\ldots \\
HCG\,30  & 04 36 28.3 & $-02$ 50 02.9 & 63 & {\em Chandra} & 2006-02-07 
        & ACIS-S VF  & 29.0 & 5.08  \\
HCG\,37  & 09 13 35.3 & $+30$ 01 25.5 & 97 & {\em Chandra} & 2005-01-13 
        & ACIS-S VF  & 17.5 & 2.30  \\
HCG\,40  & 09 38 56.7 & $-04$ 50 55.9 & 98 & {\em Chandra} & 2005-01-29 
        & ACIS-S VF  & 31.8 & 3.64  \\
 \ldots &\ldots& \dots & \ldots & \ldots & 2005-01-29& ACIS-S VF& 14.6&\ldots\\
HCG\,44  & 10 18 02.5 & $+21$ 48 50.7 & 23 & {\em Chandra} & 2002-03-14 
        & ACIS-S F   & 19.4 & 2.16  \\
\ldots  & 10 17 38.0 & $+21$ 41 17.0 & \ldots & {\em XMM} pn & 2001-05-07 
        & EFF Thick  & 8.1 & \ldots \\
\ldots  & \ldots & \ldots & \ldots & {\em XMM} m1 &\ldots & FF Thin  & 13.4 
        &\ldots \\
\ldots  & \ldots & \ldots & \ldots & {\em XMM} m2 &\ldots & FF Thin  & 13.1 
        &\ldots \\
HCG\,97 & 23 47 25.4 & $-02$ 19 45.5 & 86 & {\em Chandra} & 2005-01-14 
        & ACIS-S VF  & 36.2 & 3.65  \\
HCG\,100 & 00 01 25.9 & $+13$ 06 30.4 & 69 & {\em Chandra} & 2006-06-12 
        & ACIS-I VF  & 26.1 & 4.40  \\
 \ldots & \ldots&\ldots &\ldots&\ldots& 2007-01-24& ACIS-I VF & 16.1 &\ldots \\
\hline 
\end{tabular} 
\end{minipage} 
\end{table*}

\section{Data reduction and analysis}\label{sec,data}

\subsection{{\em XMM-Newton} data}

The {\em XMM} data were analysed using {\sc xmmsas} v6.5.0, and
calibrated event lists were generated with the `emchain' and `epchain'
tasks. Event files were filtered using standard quality flags, while
retaining only patterns $\leq 4$ for pn and $\leq 12$ for MOS.
Screening for background flares was first performed in the 10--15~keV
band for MOS and 12--14 keV for pn. Following an initial removal of
obvious large flares, a 3$\sigma$ clipping of the resulting lightcurve
was applied.  Point sources were then identified by combining the
results of a sliding-cell search (`eboxdetect') and a maximum likelihood
point spread function fitting (`emldetect'), both performed in five
separate energy bands spanning the total range 0.3--12 keV. In order
to filter out any remaining soft protons in the data, a second
lightcurve ($3\sigma$) cleaning was then done in the 0.4--10~keV~band,
within a 9--12~arcmin annulus which excluded the detected point
sources.  Closed-filter data from the calibration database and
blank-sky background data \citep{read03} for the appropriate observing
mode were filtered similarly to source data, and screened so as to
contain only periods with count rates within $1\sigma$ from the mean
of the source data.  All point sources were excised out to at least
25~arcsec in spectral analysis.
 
To aid the search for diffuse X-ray emission within the groups,
smoothed exposure-corrected images were produced, with background maps
generated from blank-sky data. We allowed for a differing contribution
from the non-vignetted particle background component in source- and
blank-sky data by adopting the following approach. First an EPIC
mosaic image was smoothed adaptively ($3\sigma$--$5\sigma$
significance range), and the particle background was subtracted. The
latter was estimated from closed-filter data which were scaled to
match source data count rates in the image energy band in regions
outside the field of view, and then smoothed at the same spatial
scales as the source data. The resulting photon image includes the
X-ray background at the source position. To remove this component, a
particle-subtracted blank-sky image was produced in a similar way, and
scaled to match 0.3--2~keV source count rates in a
point-source--excised 10--12~arcmin annulus assumed to be free of IGM
emission (this assumption is clearly justified in all cases, as will
be shown). This image was then subtracted from the corresponding
source image, and the result was finally exposure-corrected.

\subsection{{\em Chandra} data}

For all {\em Chandra} data sets, calibrated event lists were
regenerated using {\sc ciao} v3.3. For Very Faint mode observations,
the standard additional background screening was carried out. Bad
pixels were screened out using the bad pixel map provided by the
pipeline, and remaining events were grade filtered, excluding {\em
ASCA} grades 1, 5, and 7.  Periods of high background were filtered
using $3\sigma$ clipping of full--chip lightcurves, binned in time
bins of length 259.8-s and extracted in off-source regions in the
2.5--7~keV band for back-illuminated chips and 0.3--12~keV for
front-illuminated chips.  Blank-sky background data from the
calibration database were screened and filtered as for source data,
and reprojected to match the aspect solution of the latter.  Point
source searches were carried out with the {\sc ciao} task `wavdetect'
using a range of scales and detection thresholds, and results were
combined. Source extents were quantified using the $4\sigma$ detection
ellipses from `wavdetect', and these regions were masked out in all
spectral analysis.
 
In order to produce smoothed images as for the {\em XMM} data,
background maps were generated using blank-sky data, and scaled to
match source count rates for each CCD. This scaling employed either
the full-chip 10--12~keV count rates, with point sources excluded, or,
where possible, count rates in the image energy band within
source-free regions on the relevant CCD. The background maps were then
smoothed to the same spatial scales as the source data and subtracted
from the latter. The resulting images were finally exposure corrected
using similarly smoothed, spectrally weighted exposure maps, with
weights derived from spectral fits to the integrated diffuse emission
(where possible, otherwise the exposure maps were weighted by a
$T=1$~keV, $Z=0.3$~Z$_\odot$ thermal plasma model).

\subsection{Spatial and spectral analysis}\label{sec,spat}

While the smoothed X-ray images described above are useful in terms of
establishing the presence and rough morphology of any intragroup
medium, we emphasize that they were used for illustrative purposes
only and not for quantitative analyses. Where IGM emission was not
immediately obvious from these images, we performed an additional
source detection procedure based on Voronoi tessellation and
percolation (`vtpdetect' in {\sc ciao}), which can be useful for
detecting extended, low--surface brightness emission missed by our
standard detection algorithms. To reduce the fraction of spurious
detections, a minimum of 50~net~counts were required for a source to
be considered real.

As a second step in the search for group-scale diffuse emission, we
also extracted exposure-corrected 0.3--2~keV surface brightness
profiles from the unsmoothed data, with all detected point-like and
extended galactic sources masked out.  The profiles were extracted
from the optical group centre defined by the principal members in the
\citet{hick82} catalogue.  For {\em XMM} data, we used an EPIC mosaic
image for this purpose, with the particle background removed using the
method described above. The estimated particle level shows a typical
standard error of the mean of $\approx 5$~per~cent, which should be
representative of the uncertainty associated with particle subtraction
if the ratio of particle events inside and outside the field of view
is similar to that in the closed-filter data. We used the blank-sky
background data to confirm this assumption (since these have very
little source contamination), but have conservatively added a
10~per~cent error in quadrature to our {\em XMM} surface brightness
errors, to allow for any residual systematic uncertainties associated
with the particle subtraction.

For the spectral analysis of any extended emission, X-ray spectra were
accumulated in energy bins of at least 20~net counts, and fitted in
{\sc xspec} v11.3 assuming an APEC thermal plasma model with the solar
abundance table of \citet{grev98}. {\em XMM} background spectra were
extracted by means of the common 'double-subtraction' technique
\citep{arna02}, using blank-sky background data for the on-chip
background, and a large-radius (10--12~arcmin) annulus for determining
the local soft X-ray background. Owing to the smaller field of view, a
similar approach was not generally possible or desirable for the {\em
Chandra} observations where source emission may completely fill the
CCD under investigation. The extraction of {\em Chandra} background
data products are therefore described individually for each group in
the next Section.

Surface brightness profiles of the X-ray detected groups were
extracted from the peak of the diffuse X-ray emission when clearly
identifiable (in HCG\,37 and 97) and from the centroid otherwise
(HCG\,15 and 40). The profiles were fitted with standard
$\beta$--models for conversion into IGM density profiles under the
assumption of isothermality. Since X-ray emissivity is very nearly
independent of temperature for a $T\sim 0.5$--1~keV plasma of the
relevant metallicities (see \citealt{suth93}), this approach is
entirely adequate for our purposes, where the uncertainties of our
analysis will ultimately be dominated by those related to the
modelling of the {\em impact} of the hot gas on the galaxies. The
density profiles were normalized using the spectral normalization $A$
from {\sc xspec},
\begin{equation}
  A = \frac{10^{-14}}{4\pi D^2 (1+z)^2} 
  \int_{V}^{}n_e n_{\rm H} \mbox{ d}V  \mbox{ cm$^{-5}$}, 
\label{eq,xspec}
\end{equation} 
where $D$ is the assumed group distance, and $n_e$ and $n_{\rm H}$ are
the number densities of electrons and hydrogen atoms, respectively.
The integral represents the fitted emission integral $I_{\rm e}$ over
the covered volume $V$, assumed to be spherically symmetric.  Total
gas masses within the region of interest were derived by simple volume
integration of $n_e(r) \mu_e m_p$, where $m_p$ is the proton mass and
we have assumed $n_e/n_{\rm H} = 1.165$ and $\mu_e = 1.15$,
appropriate for a fully ionized $Z=0.3$~Z$_\odot$ plasma at the
relevant temperatures \citep{suth93}.

Since we have no knowledge of the density distribution of any hot gas
in the X-ray undetected groups, we have only derived constraints on
their {\em mean} electron density $\langle n_e \rangle$ within the
region considered, effectively assuming a uniform IGM
distribution. The advantage of this approach is that it provides very
conservative upper limits to the IGM masses (and mean ram pressures)
within the relevant region. The derived limits to $\langle n_e
\rangle$ and $M_{\rm IGM}$ for these groups were obtained from the IGM
count rate limits and thus depend on the depth of the X-ray data.
These count rate limits were established on the basis of the
exposure-corrected (and, in the case of {\em XMM},
particle-subtracted) images, by comparing the emission level within a
region centred on the optical group centre with that in a surrounding
annulus. The physical extent of the region of interest was thus
constrained by the need to evaluate the background locally from our
data, and varies from 50--150~kpc among the groups, as detailed in the
discussion of individual groups below. The derived constraints on IGM
count rates were translated into constraints on $I_{\rm e}$, assuming
the cooling function $\Lambda(T,Z)$ of \citet{suth93} and an IGM
temperature taken from the $\sigma_{\rm V}$--$T_{\rm X}$ relation of
\citet{osmo04},
\begin{equation} 
  \mbox{log}\, \sigma_{\rm V} = (1.15\pm0.26)\mbox{ log}\, T_{\rm X} +  
  2.60\pm 0.03,
\label{eq,sigma_T} 
\end{equation} 
with galaxy velocity dispersions $\sigma_{\rm V}$ in km~s$^{-1}$ and
$T_{\rm X}$ in keV. Errors on $T$ were derived from the dispersion of
this relation, with $\sigma_{\rm V}$ taken from P96 for HCG\,7, 30,
37, 44, and 100, from \citet{mahd05} for HCG\,97, and from
\citet{osmo04} for the remainder. The resulting temperature range was
then used to estimate $3\sigma$ upper limits on $\langle n_e \rangle
\sim (I_{\rm e}/V)^{1/2}$ inside the assumed spherical volume $V$ for
any subsolar metallicity $Z$. The assumption of a uniform IGM in these
groups implies that we can identify the upper limit to the central IGM
density $n_0$ with $\langle n_e \rangle$ and constrain the IGM masses
by simply multiplying $\langle n_e \rangle$ and $V$.

In order to briefly investigate the level of nuclear X-ray activity
among individual group members, 0.3--2~keV count rates of all galactic
central point sources were also extracted, adopting extraction regions
of 2 and 15~arcsec radius for {\em Chandra} and {\em XMM} data,
respectively. In the majority of cases, photon statistics were
insufficient to allow robust spectral fitting for individual
sources. For consistency, all point source count rates were therefore
converted to luminosities assuming a power-law spectrum of photon
index $\Gamma=1.7$, absorbed by the Galactic value of $N_{\rm H}$. The
associated uncertainties were derived from the Poisson errors on the
photon count rates.

\section{Results}\label{sec,results}
 
In this section we discuss the results obtained for the IGM in each
group and for the X-ray emission associated with individual group
galaxies. Figure~\ref{fig,mosaic} shows contours of the smoothed,
background-subtracted X-ray emission of each group overlayed on
Digitized Sky Survey (DSS) images.  Diffuse X-ray emission associated
with an intragroup medium is detected in four of the eight groups, as
described for each group individually below. For the remaining four,
we do not detect any extended group emission, neither inside a given
physical radius from the optical group centre when compared to the
emission level in surrounding regions, nor on the basis of the Voronoi
source detection procedure. To corroborate these results,
Figures~\ref{fig,surfbright} and \ref{fig,surfbright2} show the
derived surface brightness profiles for the groups with and without
detectable intragroup emission, respectively.
\begin{figure*}
  \includegraphics[width=150mm]{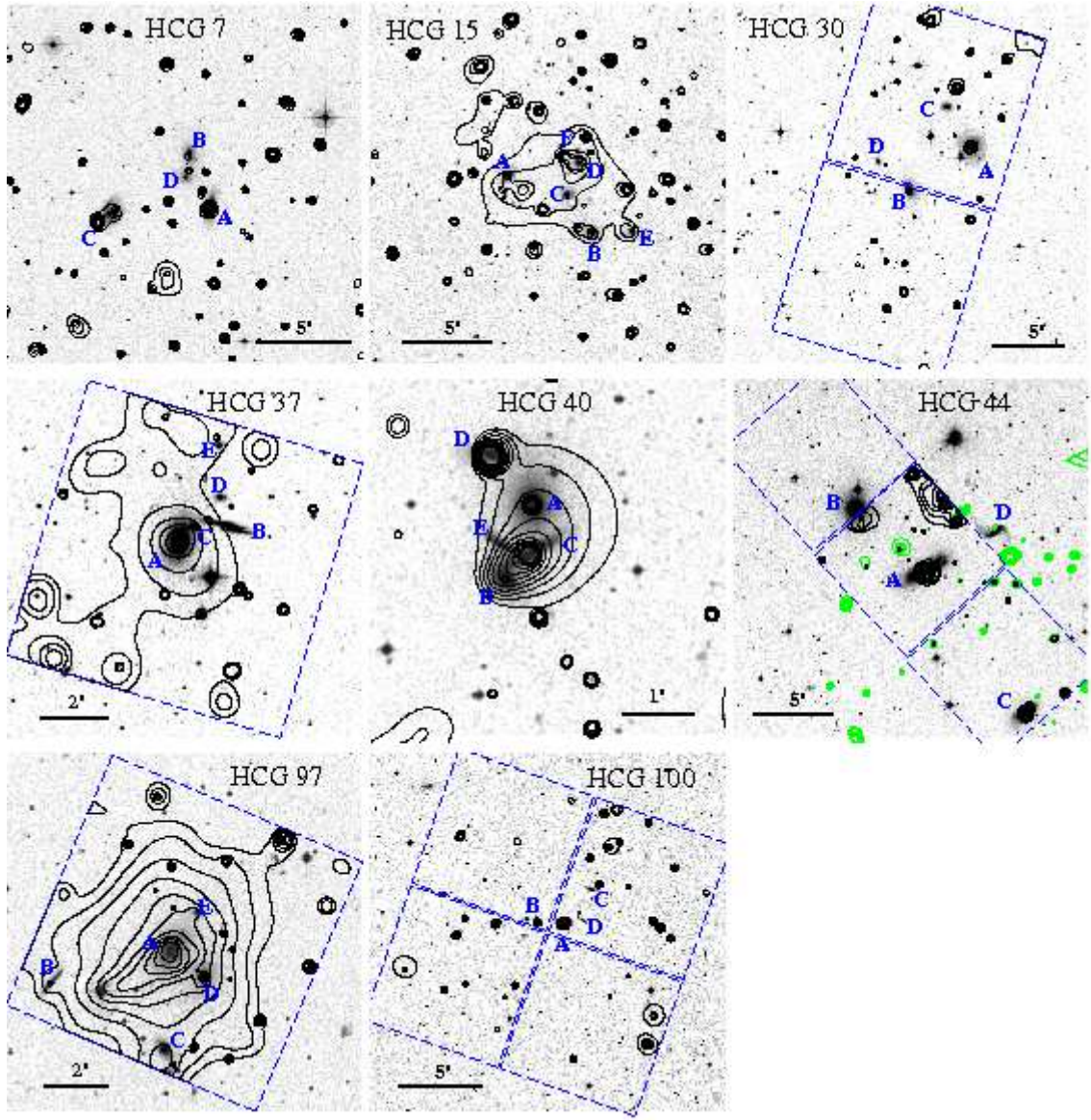}
  \caption{Contours of adaptively smoothed 0.3--2~keV emission
    overlayed on DSS images for all groups, with the principal group
    members labelled following the notation of \citet{hick82}. Where
    relevant, dashed squares outline the coverage of the {\em
    Chandra}/ACIS CCD's. For HCG\,44, dark (black) contours outline
    the {\em Chandra} emission, and lighter (green) those of an
    overlapping {\em XMM} pointing.}
\label{fig,mosaic}
\end{figure*}

Table~\ref{tab,summary} summarizes the observed H{\sc i} and X-ray
properties of the groups, along with the adopted velocity dispersions
from optical spectroscopy. H{\sc i}~deficiencies in the Table are from
our GBT measurements (Borthakur et al., in prep.), except for HCG\,44,
for which we have adopted the older VLA value \citep{verd01} due to
the GBT beam size only covering the central region of this relatively
nearby system. The listed H{\sc i} deficiencies are based on the
integrated H{\sc i} mass within the circular region covered by the
radio data ($r_{\rm HI}$ in the Table). In many groups, a significant
fraction of the detected H{\sc i} is located outside the optical
extent of individual galaxies (i.e.\ is intergalactic) and cannot be
clearly assigned to any individual group member. The derived values of
$\Delta_{\rm HI}$ should therefore generally be viewed as an average
for the galaxies within the GBT or VLA field. As the fractional
1-$\sigma$ uncertainty on the measured H{\sc i} masses from our GBT
data is less than 1~per~cent for all groups, uncertainties on the
listed deficiencies are dominated by those related to the predicted
logarithmic H{\sc i} mass, which we have taken to be 0.2, adopting the
standard estimate of error provided by \citet{hayn84}.

\begin{table*} 
 \centering 
  \begin{minipage}{122mm} 
   \caption{Summary of derived group properties. Except where
   indicated, H{\sc i}~deficiencies $\Delta_{\rm HI}$ are from GBT
   data (Borthakur et al., in prep.), obtained inside a radius $r_{\rm
   HI}$.  Column~7 lists the derived constraints on central hot IGM
   density, with upper limits for the X-ray undetected groups given at
   $3\sigma$ significance, and Column~8 the corresponding hot IGM mass
   within $r_{\rm HI}$.}
  \label{tab,summary}
  \begin{tabular}{@{}llrrcrrr@{}} \hline 
   \multicolumn{1}{l}{Group} &
\multicolumn{1}{l}{$\Delta_{\rm HI}$} &
 \multicolumn{1}{c}{$r_{\rm HI}$} &
\multicolumn{1}{c}{$\sigma_{\rm V}$} &
\multicolumn{1}{c}{$T_{\rm X}$} &
\multicolumn{1}{c}{$L_{\rm X}$} &
\multicolumn{1}{c}{$n_0$} &
\multicolumn{1}{c}{$M_{\rm IGM}$} \\ 
 & & (kpc) &  (km~s$^{-1}$) & (keV) & ($10^{40}$~erg~s$^{-1}$) 
 & ($10^{-3}$~cm$^{-3}$) & ($10^{10}$~M$_\odot$) \\ \hline
HCG\,7  & 0.60 & 68 & 95$^a$         & $0.3\pm 0.1^\ast$ & $<0.5$ &
  $<0.06$ & $<0.26$\\
HCG\,15 & 0.46 & 113 & 404$^b$     & $0.83^{+0.09}_{-0.06}$  &  
  $32\pm 2$  & $4.4\pm 0.5$   & $7.4\pm 0.8$ \\
HCG\,30 & 1.37  & 79 & 72$^a$      & $0.2\pm 0.1^\ast$ & $<1.3$ &
  $<1.2$  & $<8.3$\\
HCG\,37 & 0.33 & 119 & 446$^a$        & $0.86^{+0.13}_{-0.09}$ & $16\pm 2$ & 
  $38\pm 6$ & $6.5\pm 1.0$ \\
HCG\,40 & 0.60 & 121 & 157$^b$        & $0.59^{+0.11}_{-0.12}$ & 
  $3.1\pm 0.5$ & $1.1\pm 0.2$ & $2.4\pm 0.4$ \\
HCG\,44 & 0.69$^\dag$ & 101 & 145$^a$  & $0.4\pm 0.1^\ast$ & 
  $<3.6$ & $<0.1$ & $<1.4$ \\
HCG\,97 & 0.35 & 106 &  383$^c$ & $0.97^{+0.14}_{-0.12}$ & 
  $120^{+20}_{-24}$ & $48\pm 2$ & $13.8\pm 0.6$\\
HCG\,100& 0.27 & 86 & 100$^a$      & $0.3\pm 0.1^\ast$ & $<3.4$ & 
  $<0.33$ & $<2.9$ \\ \hline
\end{tabular}

\medskip
$^\dag$From VLA data \citep{verd01}. \\
$^\ast$Obtained from the assumed $\sigma_{\rm V}$--$T_{\rm
   X}$ relation, equation~(\ref{eq,sigma_T}).\\
$^a$\citet{ponm96}. \\
$^b$\citet{osmo04}. \\
$^c$\citet{mahd05}.
\end{minipage}
\end{table*}

For reference, 0.3--2~keV X-ray luminosities are also listed in
Table~\ref{tab,summary}, corrected for Galactic absorption, and
derived within the region employed for the spectral analysis unless
otherwise specified in the subsection for the relevant group. The
listed central hot IGM densities (or upper limits to the mean
densities for the X-ray undetected groups) were computed as outlined
in Section~\ref{sec,spat}. IGM masses in the Table were derived within
the same region as the H{\sc i}~deficiencies (i.e.\ within $r_{\rm
HI}$ given in the Table), to enable a direct comparison between the
two. Note, as indicated above, that the upper limits to $M_{\rm IGM}$
for the X-ray undetected groups conservatively assume a uniform IGM
distribution. If instead assuming a standard $\beta$--model for the
hot gas in these groups, with central density equal to the inferred
mean value $\langle n_e \rangle$ and with typical group values of,
e.g., $\beta =0.5$ and $r_c=20$~kpc, the derived IGM mass limits would
be reduced by factors of 4--7.
\begin{figure*}
\mbox{
  \includegraphics[width=75mm]{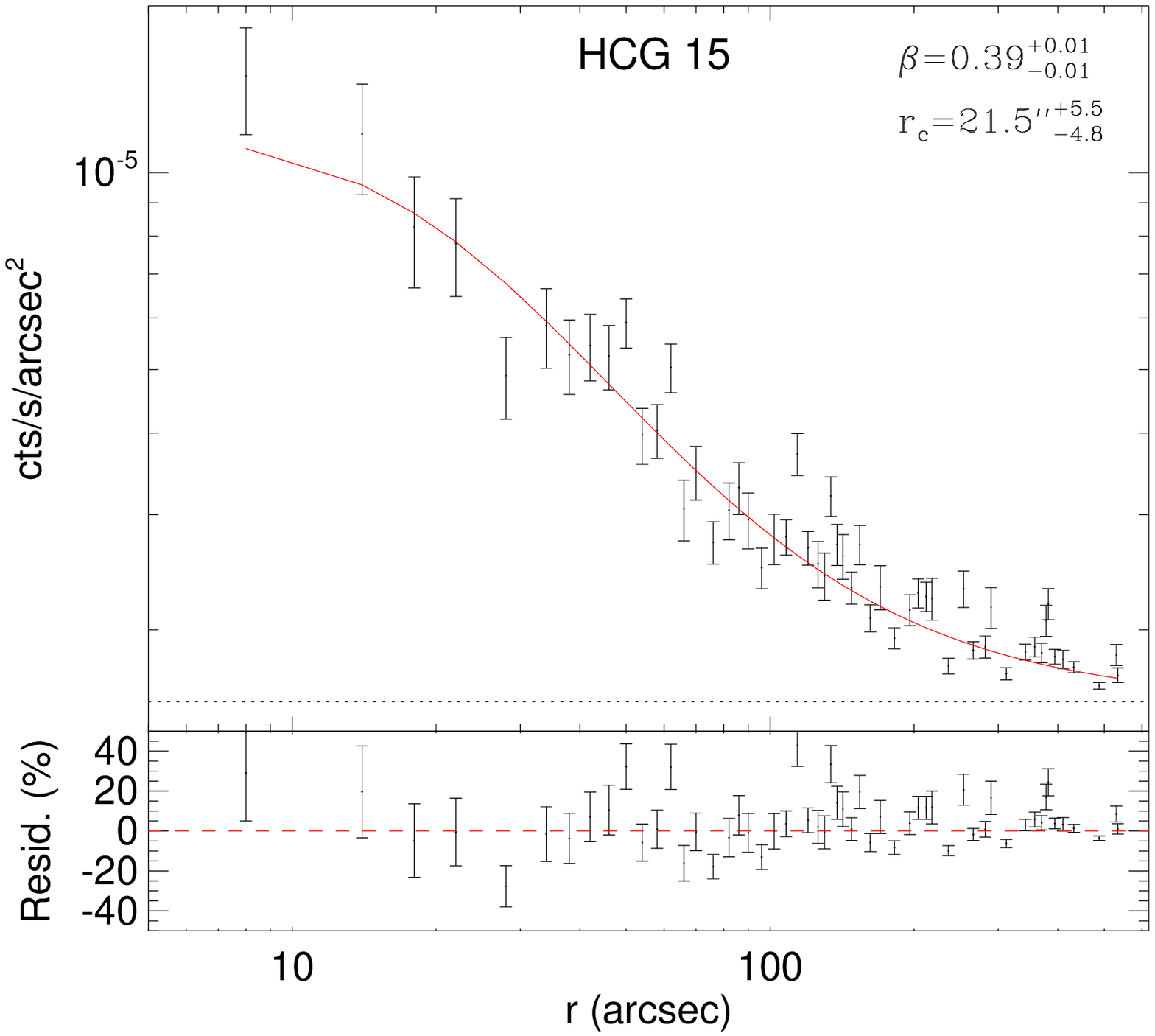}
  \includegraphics[width=75mm]{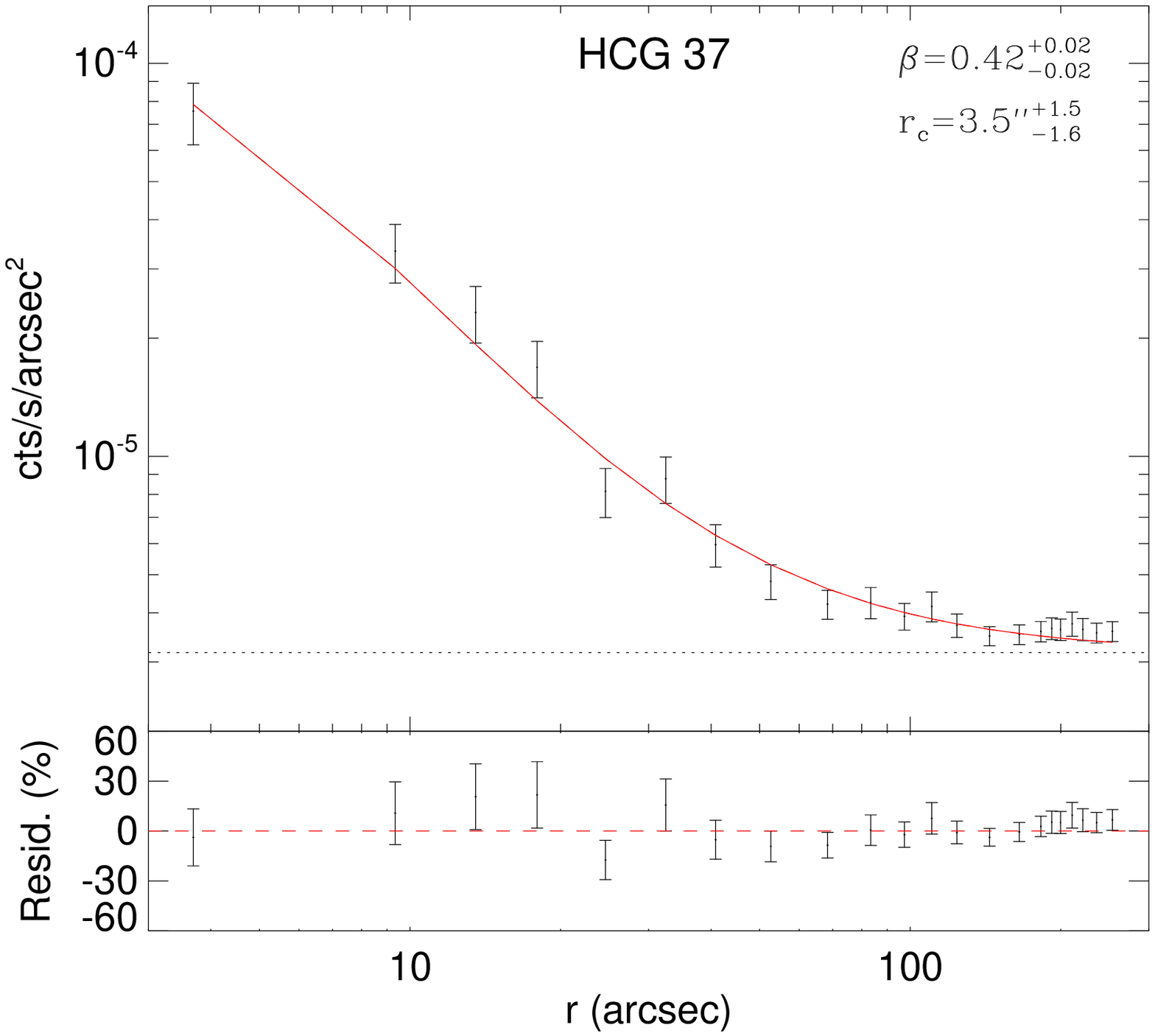}}
\mbox{
  \includegraphics[width=75mm]{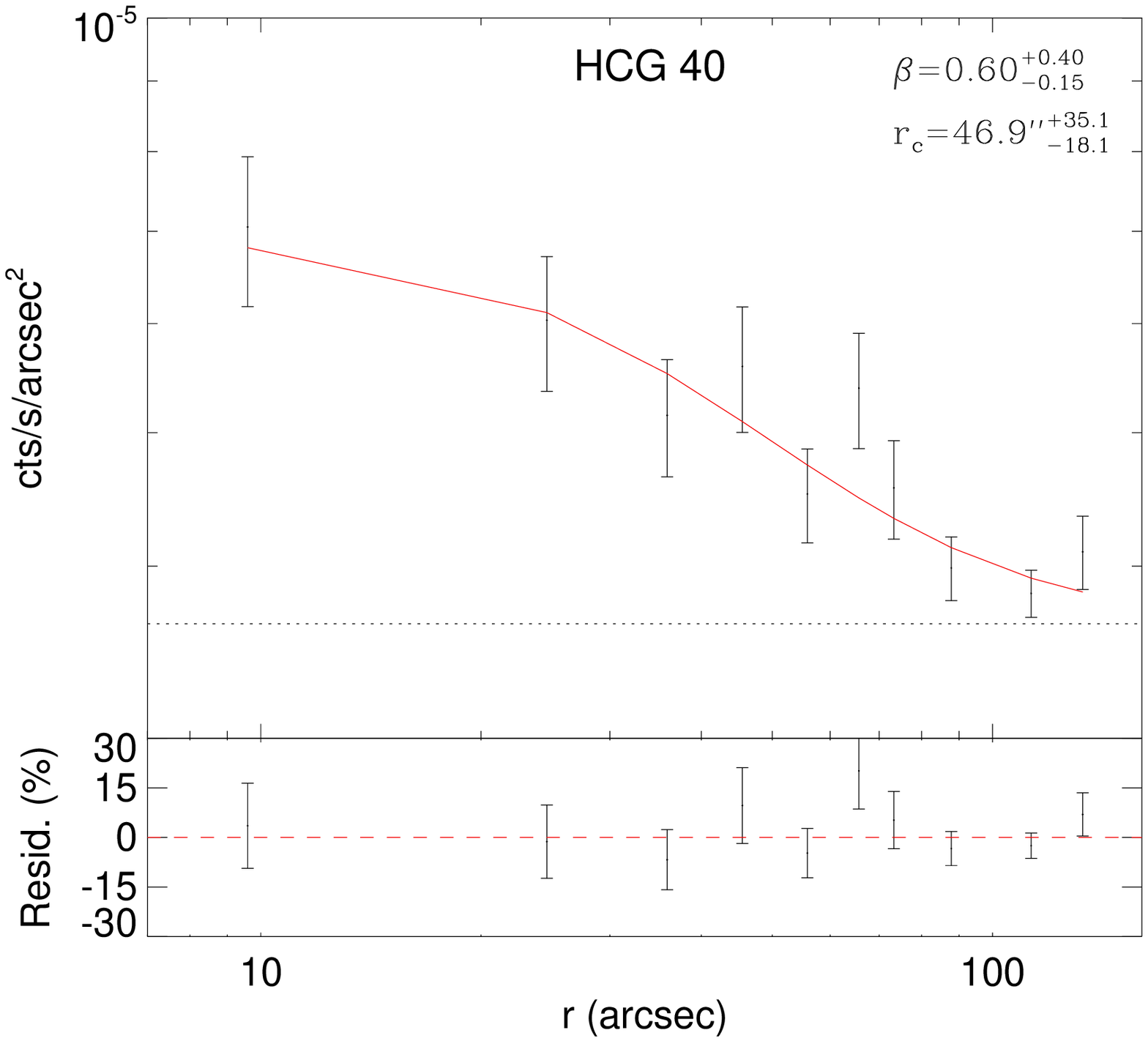}
  \includegraphics[width=75mm]{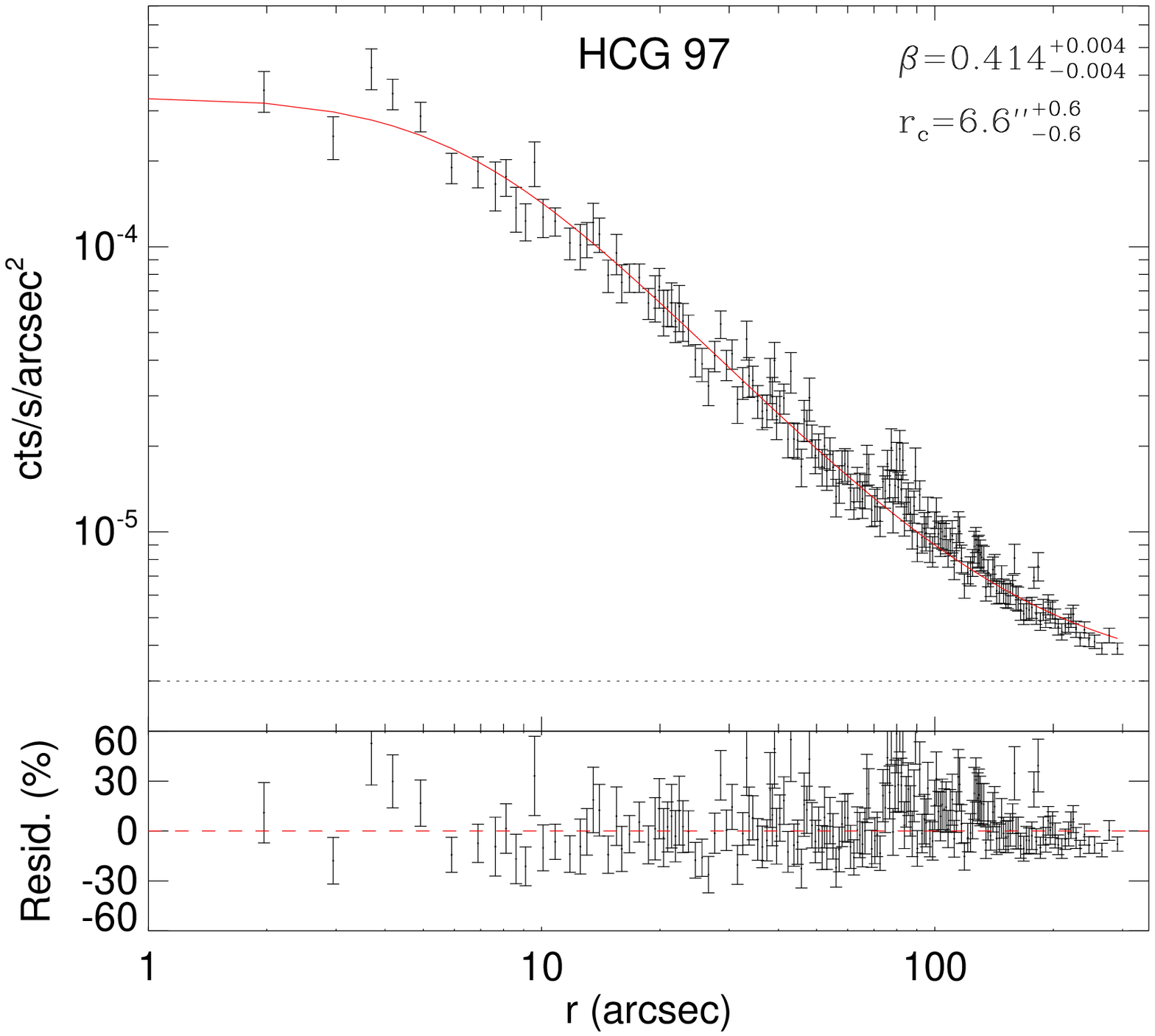}}
  \caption{0.3--2~keV exposure-corrected surface brightness profiles
   of groups with detectable diffuse emission, along with best-fitting
   $\beta$--models (solid lines). Horizontal dotted lines mark the
   estimated background level in each case. For each plot, the bottom
   panel shows fit residuals relative to the best-fitting model.}
\label{fig,surfbright}
\end{figure*}

\begin{figure}
  \includegraphics[width=80mm]{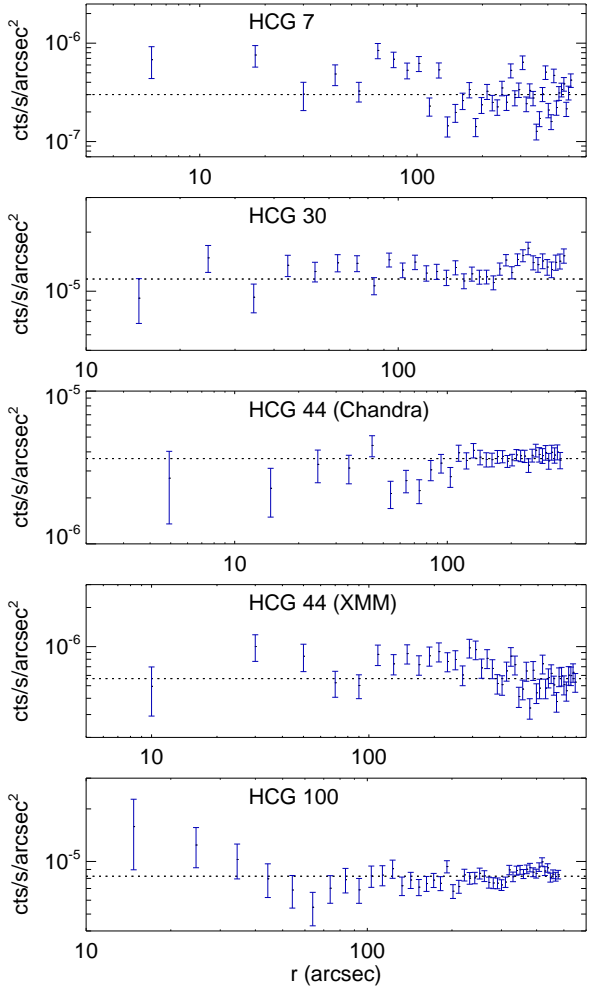}
  \caption{As Fig.~\ref{fig,surfbright}, but for the groups without
  detectable diffuse emission.}
\label{fig,surfbright2}
\end{figure}

\subsection{X-ray properties of the intragroup medium}\label{sec,IGM}

\subsubsection{HCG\,7}

This group remained X-ray undetected in shallow {\em ROSAT} All-Sky
Survey (RASS) data. Despite the {\em XMM} data of this target
representing the deepest X-ray observation within our sample, no
diffuse X-ray emission is detected in the group when comparing the
emission level of the exposure-corrected and particle-subtracted
0.3--2~keV mosaic image inside $r=9$~arcmin ($r\approx 140$~kpc) with
that measured in a surrounding annulus. This is corroborated by the
derived surface brightness profile shown in
Fig.~\ref{fig,surfbright2}. Although this profile does seem to hint at
a weak signal inside $r\approx 2$~arcmin, the combined signal inside
this region is significant at less than $1.8\sigma$, is not picked up
by `vtpdetect', and is not discernible in the smoothed image presented
in Fig.~\ref{fig,mosaic}.  Thus, we conservatively treat it as a
non-detection.

In fact, no extended emission unassociated with individual galaxies is
identified by `vtpdetect' within the 9~arcmin radius, with the
exception of the source visible in Fig.~\ref{fig,mosaic} roughly $\sim
5$~arcmin south of the optical group centre. There are no
optical/infra-red counterparts to this X-ray source listed in NED
within a 2~arcmin diameter, despite the proximity of the group
($D\approx 54$~Mpc). The emission is detected out to 2.2~arcmin from
the X-ray centroid at $3\sigma$ above the local background, confirming
that it is clearly extended. A thermal plasma model fit to the
spectrum extracted from the pn data within $r=1.5$~arcmin of the
centroid provides an acceptable fit, with a reduced
$\chi^2_{\nu}=0.92$ for 24 degrees of freedom (d.o.f.). This yields a
best-fitting temperature $T=2.6^{+0.7}_{-0.4}$~keV and redshift
$z=0.41^{+0.12}_{-0.02}$ for an assumed abundance of 0.3~solar. A
simple power-law model with Galactic absorption yields $\Gamma =
1.84\pm 0.10$ for the power-law index, but the fit is not preferred to
a thermal model (red. $\chi^2_{\nu}=1.07$ for 25 d.o.f, i.e.\ a change
of $\Delta \chi^2 = 4.8$).  The estimated temperature can be compared
to that expected for the IGM in HCG\,7 on the basis of its very low
galaxy velocity dispersion, $\sigma_{\rm V}=95$~km~s$^{-1}$, for which
the $\sigma_{\rm V}$--$T_{\rm X}$ relation of \citet{osmo04} would
suggest only $T=0.3\pm 0.1$~keV.  Combined with the redshift estimate
of $z\approx 0.4$, this strongly suggests that this emission is not
associated with HCG\,7 itself. The X-ray centroid also coincides to
within 10~arcsec with an NVSS source with a 1.4~GHz flux of 19.4 mJy
(corresponding to $1\times 10^{25}$~W~Hz$^{-1}$ at $z=0.41$), so the
X-ray emission is conceivably associated with a $z\approx 0.4$
background cluster harbouring a central radio-loud galaxy.
 
For $T$ anywhere in the range 0.2--0.4~keV and assuming any subsolar
metallicity, our failure to detect IGM emission inside $r\approx
150$~kpc translates into a $3\sigma$ upper limit to the unabsorbed
0.3--2~keV luminosity inside this region of $L_{\rm X} < 5\times
10^{39}$~erg~s$^{-1}$, with a corresponding limit to the mean gas
density of $\langle n_e\rangle <6\times 10^{-5}$~cm$^{-3}$.  We note
that HCG\,7 is included in the group catalogue of \citet{yang07}, with
a total group mass, as estimated from its optical properties, ranging
from 2.4--5.0$\times 10^{12}$~M$_\odot$ depending on the method
assumed.  This places HCG\,7 at the very low-mass end of the group
mass function, with a mass similar to that of the Local Group. Thus,
it is perhaps not surprising that we fail to detect any IGM emission
in this system.

\subsubsection{HCG\,15}

This group has a relatively high velocity dispersion of $\sigma_{\rm
V}\approx 400$~km~s$^{-1}$, and hot intragroup gas was already
detected in pointed {\em ROSAT} observations (P96).  Significant IGM
emission is seen in the {\em XMM} data presented in
Fig.~\ref{fig,mosaic}, revealing a somewhat disturbed X-ray
morphology. Despite this irregularity of the emission, an optically
bright early-type galaxy is present roughly at centre of the X-ray
emission, as is typical for fairly undisturbed X-ray bright
groups. The facts that this galaxy is a lenticular rather than an
elliptical, and that the IGM emission is not (yet) strongly peaked on
this galaxy, may suggest that the group is in the late stages of
dynamical relaxation.

Emission is detected in the imaging data out to $r=8.9$~arcmin
($r\approx 230$~kpc) at $5\sigma$ above the X-ray background level
evaluated from a surrounding annulus. A radial surface brightness
profile is shown in Fig.~\ref{fig,surfbright}, extracted from the
centroid of emission in bins containing a signal-to-noise ratio of
S/N\,$\geq 5$. As expected from the irregular X-ray morphology, a
standard $\beta$--model is not a satisfactory description of this
profile, with $\chi^2_{\nu}$ of 3.5 for 52 d.o.f. The data show
significant deviations from the best-fitting model (with
$\beta=0.39\pm 0.01$ and $r_c=21.5^{+5.5}_{-4.8}$~arcsec) at all
radii.  However, the fit residuals do not exhibit any systematic
radial variation, suggesting they are caused by local fluctuations in
the IGM distribution rather than large-scale inhomogeneities. Hence,
despite the fact that the $\beta$--model fit is clearly statistically
unacceptable, it remains useful for our purposes as a means of
characterizing the global hot gas distribution.  From inspection of
Fig.~\ref{fig,surfbright}, it is also not clear that a different, or
more complex, model would be able to provide a better description.

The relatively broad point spread function (PSF) of {\em XMM}, not
accounted for in the surface brightness fitting, could potentially
affect the observed profile at small radii, and hence the derived core
radius and central gas density. We do not expect this to be an
important effect, however, because even just the innermost radial bin
in Fig.~\ref{fig,surfbright} extends to $r=12$~arcsec, roughly twice
the full-width at half maximum of the EPIC PSF. The fact that the
best-fitting core radius is another factor of two larger also suggests
that PSF blurring does not have a significant impact on the derived
results.

Using the double-subtraction approach for extracting a background
spectrum, a fit to the global 0.3--5~keV spectrum extracted inside
$r=6$~arcmin ($r=150$~kpc) gives a temperature
$T=0.83^{+0.09}_{-0.06}$~keV and abundance $Z=0.03\pm 0.01$~Z$_\odot$,
thus confirming the low abundance derived from {\em ROSAT} data within
the same region \citep{osmo04}. These values are consistent with those
obtained using local background subtraction, but results are better
constrained due to the superior statistics of the blank-sky background
data. The derived flux and surface brightness profile imply a central
hot gas density $n_{0} = 4.4\pm 0.5 \times 10^{-3}$~cm$^{-3}$.

\subsubsection{HCG\,30}

Despite this group representing the most H{\sc i}~deficient system
within our sample, the {\em Chandra} data do not reveal any clear
evidence for diffuse IGM emission. No extended sources outside
individual galaxies are detected by `vtpdetect', thus corroborating
the RASS-based result of P96. With a galaxy velocity dispersion of
only 72~km~s$^{-1}$, the \citet{osmo04} scaling relations would
suggest a very low IGM temperature of $T= 0.2\pm 0.1$~keV. In order to
test for the presence of any such gas, we generated 0.2--0.4~keV
images of the data on the S2 and S3 CCDs separately. These images were
exposure-corrected and smoothed but not background-subtracted, in an
attempt to suppress any bias related to ACIS calibration uncertainties
at these low energies. The results reveal no clear spatial variations
in the diffuse emission on either chip, suggesting emission at a level
consistent with the local background.

As a further test, we searched the unsmoothed data for a radial
gradient in the exposure-corrected 0.3--2~keV emission level across
the S2 and S3 CCD's, finding no significant variation with distance
from the optical group centre (see Fig.~\ref{fig,surfbright2} for a
surface brightness profile extracted on the S3 CCD). This implies that
any diffuse IGM emission would have to be near-uniformly distributed
on scales of $\sim 300$~kpc, an unlikely scenario for this
low-$\sigma$ system, in which the angular extent of the region
encompassing the four principal group members is only $\approx
5$~arcmin ($\sim 90$~kpc).

The RASS 0.5--0.9~keV count rate in a 0.5--1~deg annulus centred on
the optical group centre is $8\sigma$ above the exposure-weighted mean
of the appropriate {\em Chandra} blank-sky data, suggesting a
considerable contribution from either background or (Galactic)
foreground emission at this position. The high background rate
relative to blank-sky data requires us to evaluate the background from
the source data, thus restricting the source region under
investigation on S3 to within $r \sim 3$~arcmin ($\sim 50$~kpc) of the
optical group centre. Assuming that the background in the data can be
safely evaluated from source-free regions on S3 outside this central
region (an assumption supported by Fig.~\ref{fig,surfbright2}), the
density of any hot IGM in the group can be constrained. Inside this
region, and for $T$ in the range 0.1--0.3~keV and any subsolar
metallicity, the $3\sigma$ upper limit to the mean gas density is
$\langle n_e \rangle < 1.2\times 10^{-3}$~cm$^{-3}$. The possibility
that $T$ is very low in this group propagates into a relatively weak
constraint on $\langle n_e \rangle$.

\subsubsection{HCG\,37}

Irregular diffuse X-ray emission was detected in this group with {\em
ROSAT} out to $r\sim 8$~arcmin \citep{mulc03}, well beyond the region
covered by a single ACIS chip in our {\em Chandra} data. The latter
clearly indicate that the group emission is sharply peaked on the
early-type galaxy HCG\,37a ($=$~NGC\,2783), the nucleus of which is
also detected as a point-like source in the data. The association of
the IGM X-ray peak with HCG\,37a was not obvious from the earlier {\em
ROSAT} data, as the much broader {\em ROSAT} PSF required
\citet{mulc03} to exclude point-like emission out to $r=1.5$~arcmin
from the peak, thus effectively masking out HCG\,37a in the
data. Despite the overall irregularity of the group emission, the
X-ray centroid (when masking out the HCG\,37a nucleus) coincides to
within 10~arcsec with the optical position of HCG\,37a as listed in
NED.

Since group emission covers the S3 CCD, we cannot reliably use any
method relying on source-free regions on S3 to evaluate the background
in the {\em Chandra} data. The situation is aggravated by the fact
that RASS data indicate a $3\sigma$ soft background deficit at this
position relative to the appropriate blank-sky data, so background
subtraction by means of these is not straightforward either. To
circumvent these issues, we adopted the method employed by
\citet{vikh05}. First, a source minus blank-sky spectrum was extracted
on the back-illuminated S1 chip, to quantify the difference in the
soft background between source- and blank-sky data. The spectrum was
fitted with a $T=0.18$~keV $Z=\mbox{Z}_\odot$ mekal plasma in the
0.4--1~keV range, with the normalization allowed to be negative. The
best-fitting model was then added to the model fit of the blank-sky
subtracted source emission on S3 inside $r=3.7$~arcmin ($r\approx
100$~kpc) after scaling to the source region area. The resulting
background level was also used for the surface brightness analysis. We
note that, at 90~per~cent confidence, the best-fitting $T$ and $Z$
resulting from this approach, $T=0.86^{+0.13}_{-0.09}$~keV and
$Z=0.10^{+0.08}_{-0.04}$~Z$_\odot$, are just consistent with the P96
values inside $r=150$~kpc ($T=0.67\pm 0.11$~keV, $Z=0.17\pm
0.15$~Z$_\odot$), lending some credibility to this approach.

The surface brightness profile shown in Fig.~\ref{fig,surfbright}
confirms the presence of emission across the full S3 CCD in the {\em
Chandra} data. The profile was centred on the X-ray peak and extracted
in bins containing at least 30~net~counts.  Despite the irregularity
of the emission on large scales (Fig.~\ref{fig,mosaic}), a
$\beta$--model provides a good fit to the profile across the full
radial range plotted in Fig.~\ref{fig,surfbright}, yielding
$\beta=0.42\pm 0.02$ and $r_c=3.5^{+1.5}_{-1.6}$~arcsec, with
$\chi^2_{\nu}=0.80$ for 19 d.o.f.  The spatial and spectral results
imply a gas density in the group core of $0.038\pm 0.006$~cm$^{-3}$.

\subsubsection{HCG\,40}

The {\em Chandra} observation of this group was split into two
separate pointings, so a merged event file was produced for the
imaging analysis. Although undetected in a 3.6-ks {\em ROSAT} pointing
(P96), Fig.~\ref{fig,mosaic} suggests the presence of diffuse emission
in this group.  The centroid of this emission, with the optical extent
of the individual group members masked out, is located $\sim
0.5$~arcmin to the NW of HCG\,40c and so is not clearly associated
with any individual galaxy.  To test that the emission seen in
Fig.~\ref{fig,mosaic} is truly extended and not simply due to the
smoothing of point sources, a surface brightness profile of the
unsmoothed, exposure-corrected emission was extracted from the
centroid in bins of at least 30~net~counts, with individual galaxies
masked out. The result is shown in Fig.~\ref{fig,surfbright}, with the
background evaluated from off--source regions on the S3 CCD. Emission
is detected above this background out to $r=2.3$~arcmin ($r=65$~kpc),
suggesting group-scale extended emission, although the detection is
only significant at $>3\sigma$ for the innermost 30~kpc. A
$\beta$--model provides an acceptable fit to this profile, with
$\chi^2_{\nu}=1.01$ for 7 d.o.f., yielding
$\beta=0.60_{-0.15}^{+0.40}$ and $r_c=46.9_{-18.1}^{+35.8}$~arcsec
($23^{+17}_{-9}$~kpc), in accordance with expectations for a typical
X-ray bright group.

For the spectral analysis of this emission, spectra and response
products were extracted separately for each of the two observations.
The spectra were then jointly fitted within the central $r=1$~arcmin,
within which the signal allows useful constraints to be obtained,
using a surrounding $2.5$--$3.5$~arcmin annulus for background
estimation. With only $\sim 140$~net~counts, the IGM abundance remains
unconstrained. Fixing $Z$ at 0.3 solar yields
$T=0.59^{+0.11}_{-0.12}$, and $T$ remains consistent with this for any
subsolar $Z$. For these parameters, the observed flux translates into
a central electron density of $1.1\pm 0.2 \times 10^{-3}$~cm$^{-3}$
and implies a diffuse 0.3--2~keV luminosity inside $r=2.3$~arcmin of
$3.1\pm 0.5\times 10^{40}$~erg~s$^{-1}$.

The extent of the emission, coupled with the fact that it is not
clearly centred on any group member, suggests that the emission is not
due to, for example, hot gas associated with an elliptical but rather
reflects the presence of a hot IGM. This interpretation would place
HCG\,40 among the relatively rare examples of spiral-dominated groups
showing intergalactic hot gas; within Hickson's (1982) catalogue, only
HCG\,16, 57, and the well-studied HCG\,92 (Stephan's Quintet) share
similar features (e.g., \citealt{dos99,fuka02,trin03}). Based on the
$B$-band luminosities of the group members, and on the $L_{\rm
X}$--$L_B$ relations for ellipticals and normal star-forming spirals
from \citet*{osul03} and \citet{read01} respectively, one would expect
a total galactic diffuse $L_X \approx 9\times 10^{40}$~erg~s$^{-1}$ in
the group, a factor of three larger than that found here for the
intragroup emission. Although care has been taken in masking out
emission from the group members, the low S/N and the compactness of
the galaxy configuration implies that we cannot exclude a residual
contribution to the diffuse emission from individual galaxies. A
conservative approach would be to regard the association of the
observed diffuse emission with an intragroup medium in HCG\,40 as
tentative rather than conclusive.

\subsubsection{HCG\,44}

While Fig.~\ref{fig,mosaic} does not indicate the presence of any IGM
emission in this system, this could simply be an artefact of the
proximity of the group ($D\approx 23$~Mpc) in combination with the
limited {\em Chandra} angular coverage, which furthermore renders
quantitative analysis of the background level in the data
non-trivial. Using local background subtraction could potentially
produce unreliable results, as IGM emission might cover the entire
ACIS array. The situation is further complicated by an enhanced
particle level in the cleaned data compared to the blank-sky files,
with the 10--12~keV count rate on the S3 CCD being 35~per~cent higher
than the corresponding blank-sky value. In addition, soft Galactic
0.5--0.9~keV emission at a level of $2\sigma$ above the blank-sky data
is also present at this position, so it is not obvious that blank-sky
data would be an appropriate choice for background estimates.

Fortunately, the presence of an overlapping {\em XMM} pointing allows
an independent test for the presence of diffuse emission in the
group. The {\em Chandra} and {\em XMM} surface brightness profiles
shown in Fig.~\ref{fig,surfbright2} suggest no detectable IGM emission
close to the optical group centre. Note that the extraction of these
profiles excluded different position angles due to the optical group
centre being close to the southern (northern) edge of the S3 (EPIC)
CCDs, so the profiles extend in largely opposite directions on the
sky. The {\em XMM} profile shows no systematic variation out to
$r=15$~arcmin ($r\approx 100$~kpc), remaining largely consistent with
the background level evaluated outside this region in the source
data. Consequently, the background level for the {\em Chandra} profile
was estimated from the northern corners of the S3 CCD, the result
suggesting no excess diffuse emission extending northwards either.

Furthermore, no extended sources that can be unambiguously associated
with group emission were detected by `vtpdetect'. In addition to the
two group galaxies HCG\,44a and b, a third extended X-ray source is
seen on the S3 chip, clearly visible in Fig.~\ref{fig,mosaic} roughly
four arcmin north of the spiral HCG\,44a, and also seen in both the
{\em XMM} data and in pointed {\em ROSAT} observations. The X-ray peak
of this source coincides with a 2MASS source with $m_K= 14.31$, but
there is no optical counterpart or redshift information available in
NED. If this source were at the group distance, the resulting $K$-band
luminosity of $2.1\times 10^8$~L$_{\odot,K}$ would place it at the
extreme faint end of the dwarf galaxy luminosity function, with the
{\em ROSAT} flux implying a ratio $L_{\rm X}/L_K \approx 0.013$, two
orders of magnitude above typical values seen even for dwarf {\em
starburst} galaxies \citep*{rasm04}. Spectral fit results provide
further support for the idea that this source is unlikely to be
associated with HCG\,44. A thermal plasma model fixed at the group
redshift returns an unacceptable fit ($\chi^2_\nu=1.61$), whereas a
significant fit improvement results when leaving $z$ as a free
parameter, yielding $\chi^2_\nu=1.37$ for 6 d.o.f, with
$T=1.98^{+0.67}_{-0.39}$ and $z=0.24^{+0.09}_{-0.07}$ for an assumed
abundance of $Z=0.3$~Z$_\odot$.

While the {\em Chandra} data are useful in terms of investigating
evidence for hot gas being stripped from individual galaxies in this
very nearby system, it is not clear that these data enable significant
improvements on the hot IGM constraints over the existing 4.7-ks {\em
ROSAT} pointing (with its much larger field of view enabling a more
reliable background subtraction in this case). Even the {\em XMM} data
can only probe emission from a quarter of the volume inside
$r=100$~kpc from the optical group centre, due to the latter being
close to the edge of the {\em XMM} field of view. Using the adopted
$\sigma$--$T$ relation, which would suggest $T=0.4\pm 0.1$~keV, the
{\em ROSAT} constraint of P96 on the X-ray luminosity inside
$r=150$~kpc ($L_{\rm X}< 3.6\times 10^{40}$~erg~s$^{-1}$ for our
adopted distance) translates into $\langle n_e\rangle < 1.0\times
10^{-4}$~cm$^{-3}$ for any subsolar metallicity. The corresponding
{\em XMM} constraint of $\langle n_e\rangle < 1.4\times
10^{-4}$~cm$^{-3}$ applies within a volume less than 10~per~cent of
that probed by {\em ROSAT}, so we have adopted the stronger {\em
ROSAT} limit in Table~\ref{tab,summary}.

\subsubsection{HCG\,97}

This constitutes the most X-ray luminous system within the sample.  A
two-dimensional analysis of a 21-ks {\em XMM} observation of this
group was performed by \citet{mahd05}, along with optical spectroscopy
identifying 37 members. Their {\em XMM} data show a plume stretching
to the southeast, beyond the region covered by
Fig.~\ref{fig,mosaic}. \citet{mahd05} speculate that this plume could
represent gas stripped from one of the central galaxies, but the {\em
XMM} data alone cannot establish this, and the {\em Chandra} data on
the S2 CCD cannot improve on the situation beyond confirming the
presence and overall morphology of this feature. One of the
spectroscopically identified member galaxies is located within the
plume but is not itself detected in either the {\em XMM} or {\em
Chandra} data. On the S3 CCD, the emission appears fairly regular in
Fig.~\ref{fig,mosaic}, albeit with the central X-ray contours somewhat
elongated towards the south-east. If masking out this elongated
feature, the centroid of the IGM emission coincides to within
10~arcsec with the position of the optically brightest group galaxy,
HCG\,97a, as listed in NED.

As established already by {\em ROSAT} observations (e,g.,
\citealt{mulc03}), diffuse emission in this group extends well beyond
the region covered by the S3 CCD, so blank-sky data were used to
evaluate the background for the {\em Chandra} surface brightness
analysis (the RASS 0.5--0.9~keV background count rate at this position
is in good agreement with the corresponding exposure-weighted mean
value of the blank-sky data).  We note, however, that the {\em
Chandra} observation was somewhat affected by background flares,
reducing the useful exposure time from 57.9 to 36.3-ks. As for
HCG\,44, the background remains high after cleaning, with the
10--12~keV count rate on S3 again being 35~per~cent above the
blank-sky value. While bearing this issue in mind, results indicate
that emission is detected above $5\sigma$ significance everywhere on
the S3 chip. A fit to the exposure-corrected 0.3--2~keV surface
brightness profile, extracted from the X-ray peak and shown in
Fig.~\ref{fig,surfbright}, yields $\beta=0.414 \pm 0.004$ and $r_c=
6.6\pm 0.6$~arcsec, with $\chi^2_{\nu} = 1.45$.  However, while
Figure~\ref{fig,surfbright} and the excellent agreement of these
results with the best-fitting parameters of \citet{mulc03} (who find
$\beta=0.41\pm 0.01$ and $r_c < 0.1$~arcmin) suggest that our
background estimate is not seriously in error, we will nevertheless
base our normalization of the density profile on the {\em ROSAT}
results of \citet{mulc03}, given the concern about the elevated
particle background in the {\em Chandra} data. Combining their
spectral results and X-ray luminosity (0.3--2~keV $L_{\rm X} =
1.20^{+0.20}_{-0.24}\times 10^{42}$~erg~s$^{-1}$ for our adopted
distance) with our surface brightness fit then implies a temperature
$T=0.97^{+0.14}_{-0.12}$~keV and central density of $0.048\pm
0.002$~cm$^{-3}$, which are the values listed in
Table~\ref{tab,summary}.

\subsubsection{HCG\,100}\label{sec,H100}

This is a group in which the H{\sc i} is clearly being stripped from
the galaxies at present, with much of it pulled into a 100~kpc long
tidal tail extending to the southwest from the optical group centre
(Borthakur et al., in prep.). In addition, the VLA data reveal a
striking H{\sc i} trail extending to the east away from the group
core, protruding from one of the galaxies in the field, Mrk\,935. This
galaxy is not included in the original \citet{hick82} catalogue, but
is a group member on the basis of its projected distance from the
optical group centre (6.7~arcmin $\sim 130$~kpc) and small radial
velocity difference of $\Delta v \sim 250$~km~s$^{-1}$ relative to the
group mean, as listed in NED. The associated H{\sc i} feature may
therefore indicate ongoing stripping as the galaxy falls into the
group. HCG\,100 thus constitutes an excellent laboratory for the study
of the processes whereby H{\sc i} is removed from individual galaxies
and heated. Unlike the case for the other {\em Chandra} observations
presented here, this group was observed using the ACIS-I array, to
allow the observed field to fully encompass all of the interesting
H{\sc i} features mentioned above.

The {\em Chandra} observation was split into two, so the imaging
analysis proceeded as for HCG\,40. The combined imaging data, shown in
Fig.~\ref{fig,mosaic}, and the resulting surface brightness profile in
Fig.~\ref{fig,surfbright2}, do not reveal any clear indications of
diffuse emission above the background level as evaluated outside
$r=8$~arcmin from the corners of the ACIS-I array. Although
Fig.~\ref{fig,surfbright2} indicates a mild net excess in the two
innermost bins, the signal within this region is significant at less
than $1.5\sigma$.  We also note that no extended sources are detected
outside individual galaxies with `vtpdetect', and that the group also
remained undetected in RASS data (P96). With the $\sigma$--$T$
relation suggesting $T=0.3\pm 0.1$~keV, the absence of an IGM
detection inside $r\approx 100$~kpc ($r\approx 5$~arcmin) from the
optical group centre implies a $3\sigma$ upper limit on the mean IGM
density of $\langle n_e \rangle < 3.3\times 10^{-4}$~cm$^{-3}$.

\subsection{Individual galaxies}\label{sec,gals}

The results presented so far demonstrate the presence of a detectable
hot IGM within half of our sample only.  However, even in the absence
of such gas, there could still be an intragroup medium present with
temperature or density below our detection limits (including any H{\sc
i} already stripped from individual galaxies, as evidenced by the GBT
detection of intergalactic H{\sc i} in many of our groups).

To explore the possibility that group galaxies could be interacting
with such a medium, and to search for signs of galaxies being stripped
of any {\em hot} gas, we present in Fig.~\ref{fig,gals} a collage of
all group members which were clearly identified as X-ray extended
sources by our source detection algorithms. These images were
adaptively smoothed following the procedure outlined in
Section~\ref{sec,data}. Exceptions are HCG\,30a and Mrk\,935, for
which a simple Gaussian smoothing (with $\sigma=10$~arcsec) was
employed due to the very low S/N. For the groups observed by {\em XMM}
(HCG\,7 and 15), the combination of group distance and the broader
EPIC PSF does not enable a clear distinction between point-like and
diffuse emission, so none of the relevant galaxies has been included
in this figure. HCG\,44b, lying on a chip gap in the {\em Chandra}
data and not covered by the overlapping {\em XMM} pointing, has also
been excluded.

\begin{figure}
\mbox{\hspace{2mm}
  \includegraphics[width=80mm]{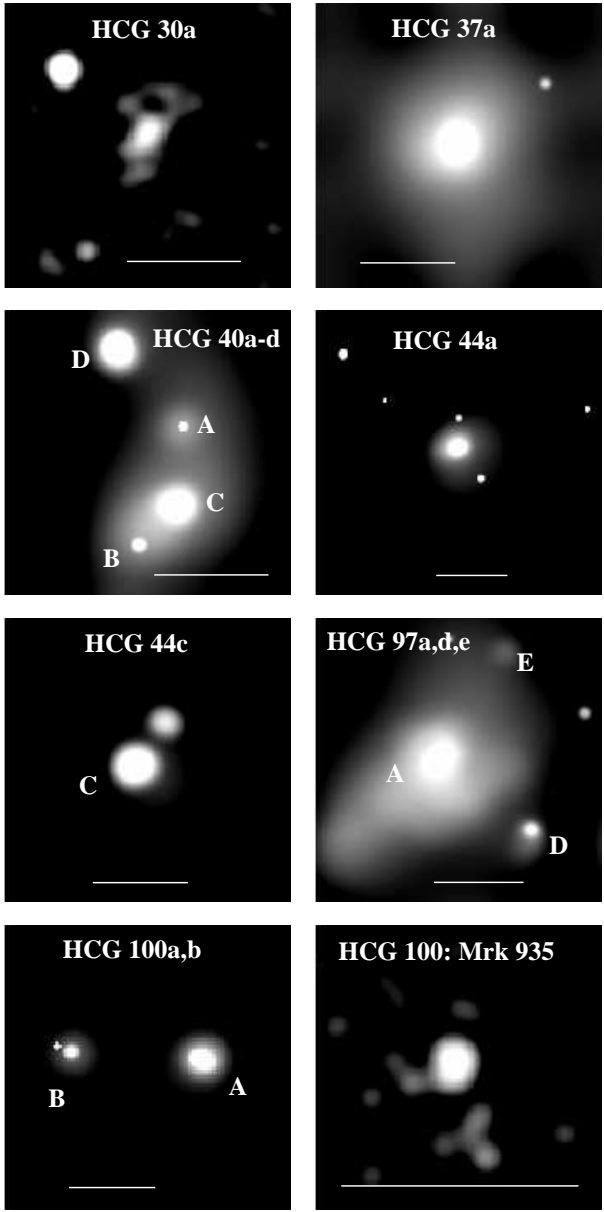}}
  \caption{Smoothed 0.3--2~keV images of individual group galaxies
  showing diffuse X-ray emission. A horizontal bar marks a scale of
  1~arcmin in each case.}
\label{fig,gals}
\end{figure}

The figure does not reveal any clear evidence for galaxies currently
being stripped of any hot gas. In particular, there are no indications
of X-ray tails or bow-shock features indicating interactions with a
surrounding medium. Such tails have been observed extending from
galaxies in clusters (e.g., \citealt*{wang04,sun05}) but seem to be
very rare in groups, with perhaps NGC\,6872 and NGC\,2276 the most
prominent examples \citep{mach05,rasm06}. There is an indication of an
asymmetric structure in HCG\,97d even in the unsmoothed data, with a
hint of a tail pointing south, but the signal is too weak to exclude
contamination by a faint point source. Mrk\,935, with the remarkable
H{\sc i} tail extending to the east, also presents evidence of some
X-ray asymmetry in this direction, but the S/N is again too low to
enable firm conclusions.

In addition to diffuse galactic emission, X-ray point sources in
individual galaxies are detected in all groups, with some of these
sources clearly associated with the galaxy nuclei. The relevance of
this is linked to the possibility that galaxies suffering strong
(tidal) interactions could be showing enhanced nuclear activity, for
example associated with a nuclear starburst or strong AGN accretion
fueled by a tidally induced gas inflow. For reference,
Table~\ref{tab,gals} lists detected X-ray sources whose position
coincides with the optical centre of individual group members, with
luminosities of any point-like component derived as described in
Section~\ref{sec,data}. Note that in some cases, such as HCG\,97e and
most of the {\em XMM} sources, we cannot clearly distinguish between
nuclear and galaxy-wide diffuse emission, and the classification of
these is followed by a `?' in the Table~\ref{tab,gals}. For HCG\,7,
however, the tentative identification of nuclear X-ray activity in
three of the four principal galaxies agrees perfectly with the {\em
Spitzer} far-infrared results of \citet{gall08}, suggesting our
identification is reasonably robust.

Among the principal members in Hickson's (1982) catalogue, we detect
22 candidate nuclear X-ray sources in 40 galaxies, with roughly
two-thirds of these falling in the groups with a detectable IGM. The
corresponding nuclear source fractions of $64\pm 17$ and $44\pm
16$~per~cent in the groups with and without detectable hot gas,
respectively, are statistically indistinguishable at the $1\sigma$
level. If instead splitting the sample according to H{\sc i}
deficiency, the corresponding fractions are $42\pm 15$ (high
$\Delta_{\rm HI}$) and $67 \pm 18$~per~cent, a difference which is
only just significant at $1\sigma$.  The median nuclear X-ray
luminosities for the two subsamples are also very similar, $4.4\times
10^{39}$ (high $\Delta_{\rm HI}$) and $4.9\times
10^{39}$~erg~s$^{-1}$, suggesting that the above conclusions are not
strongly biased by a systematic difference in limiting X-ray flux
between high-- and low-$\Delta_{\rm HI}$ groups.  We also note that
results from optical spectroscopy indicate that $\sim 40$~per~cent of
the principal members in HCGs in general show evidence for AGN
activity, with a total of $\sim 70$~per~cent showing emission lines
from either AGN or star formation activity \citep{mart07}. Our derived
fractions are generally bracketed by these values, suggesting that our
results give a reasonably reliable picture of the frequency of nuclear
activity within our sample.

With the limited statistics available, there is thus no strong
evidence from the X-ray data alone for enhanced nuclear activity
within a certain kind of groups in our sample. Specifically, if
interpreting the X-ray bright or highly H{\sc i}~deficient systems as
dynamically more evolved, we find no clear indication that the
frequency or strength of nuclear X-ray activity depends on the
dynamical status of the group. However, we note that this result
applies to a small sample and to the principal members only; a
complete census of galaxy membership from optical spectroscopy would
be required to extend this analysis to optically fainter group members
and place this conclusion on a more robust basis. The tentative lack
of a clear enhancement in nuclear X-ray activity among the most H{\sc
i}~deficient groups within our sample may tie in with the observation
that star formation activity is not globally enhanced in HCG galaxies
compared to isolated ones \citep{verd98}, as discussed in more detail
in Section~\ref{sec,discus}.

\begin{table} 
 \centering
   \caption{Overview of X-ray sources centred on individual group
     galaxies as identified by our detection algorithms. Galaxy
     morphologies were taken from NED. Column~4 lists the unabsorbed
     0.3--2~keV luminosity of any nuclear component.}
  \label{tab,gals} 
  \begin{tabular}{@{}lllc@{}} \hline 
   \multicolumn{1}{l}{Galaxy} & 
   \multicolumn{1}{l}{Morph.} & 
   \multicolumn{1}{l}{Source} &
   \multicolumn{1}{c}{$L_{\rm X,nucl}$} \\ 
  & & & (erg~s$^{-1}$) \\ \hline
HCG\,7a     & Sa   & Nuclear?  & $3.1\pm 0.1 \times 10^{40}$ \\
HCG\,7b     & SB0 & Nuclear?  & $3.8\pm 0.4 \times 10^{39}$ \\
HCG\,7c     & SBc & Nuclear   & $4.9\pm 0.4 \times 10^{39}$ \\
HCG\,15a   & S0  & Nuclear? & $2.6\pm 0.2 \times 10^{40}$ \\
HCG\,15b   & S0  & Nuclear? & $2.0\pm 0.2 \times 10^{40}$ \\
HCG\,15d   & S0  & Nuclear? & $6.7\pm 0.1 \times 10^{41}$ \\
HCG\,15e   & S0  & Nuclear? & $1.9\pm 0.2 \times 10^{40}$ \\
HCG\,30a   & SB0 & Diffuse   & -- \\
HCG\,30b   & SB0/a & Nuclear  & $3.0\pm 0.4 \times 10^{40}$ \\
HCG\,37a   & S0/E7 & Diffuse + nuclear & $1.4\pm 0.2 \times 10^{40}$ \\
HCG\,37b   &  Sbc & Nuclear  & $0.8\pm 0.4 \times 10^{39}$ \\
HCG\,40a   &  E  & Nuclear  & $2.4\pm 0.5 \times 10^{39}$ \\
HCG\,40b   &  S0 & Nuclear  & $1.8\pm 0.4 \times 10^{39}$ \\
HCG\,40c   & SBb & Diffuse?  & -- \\
HCG\,40d  & SBa & Diffuse? + nuclear & $9.0\pm 0.9 \times 10^{39}$ \\
HCG\,44a  & Sa & Diffuse + nuclear  & $1.8\pm 0.2 \times 10^{39}$ \\
HCG\,44b   & E  & Diffuse   & -- \\
HCG\,44c   & SBa & Diffuse   & -- \\
HCG\,97a  & SB0 & Diffuse + nuclear  & $7.0\pm 0.8 \times 10^{39}$ \\
HCG\,97b   &  Sc & Nuclear  & $1.8\pm 0.4 \times 10^{39}$ \\
HCG\,97c   &  Sa & Nuclear  & $0.9\pm 0.3 \times 10^{39}$ \\
HCG\,97d  & E & Diffuse? + nuclear & $8.9\pm 0.9 \times 10^{39}$ \\
HCG\,97e  & S0a & Diffuse or nuclear & $0.9\pm 0.3 \times 10^{39}$ \\
HCG\,100a &  S0/a & Diffuse + nuclear & $2.7\pm 0.5 \times 10^{39}$ \\
HCG\,100b &  S0/a & Diffuse + nuclear & $1.0\pm 0.3 \times 10^{39}$ \\
Mrk\,935  & S? & Diffuse or nuclear & $1.6\pm 0.4 \times 10^{39}$ \\
\hline 
\end{tabular} 
\end{table}

\subsection{H{\sc i}~deficiency and IGM properties}

The observed diversity in the diffuse X-ray properties of these highly
H{\sc i}~deficient groups immediately suggests that galaxy--IGM
interactions are not the dominant mechanism for driving cold gas out
of the galaxies within our sample and establishing the observed H{\sc
i}~deficiencies.  Of course, this conclusion neglects the fact that we
are not uniformly sensitive to the presence of hot gas in the
different groups.  A quantitative comparison of observed H{\sc i}
deficiencies and derived IGM properties is therefore presented in
Figure~\ref{fig,hi_def}.  The left panel shows $\Delta_{\rm HI}$ and
hot IGM mass as listed in Table~\ref{tab,summary}, with both
quantities derived within the same region (inside $r_{\rm HI}$ in the
Table). Even when considering the X-ray detected systems alone, the
obvious lack of a positive correlation between the two quantities
immediately suggests that the amount of hot gas in the group core is
not a pivotal factor for H{\sc i} removal. We note that an identical
conclusion is reached if replacing $\Delta_{\rm HI}$ with `missing'
H{\sc i} mass in the plot. The strong $3\sigma$ upper limit ($M <
2.6\times 10^9$~M$_\odot$) on the IGM mass in the highly H{\sc
i}~deficient HCG\,7 (with the largest `missing' H{\sc i} mass in the
sample of $\sim 1.8\times 10^{10}$~M$_\odot$) only reinforces this
conclusion. Also note, as pointed out in Section~\ref{sec,IGM}, that
more realistic assumptions about the IGM distribution in the X-ray
undetected groups could reduce their upper limits to $M_{\rm IGM}$ by
perhaps an order of magnitude, but that this has no bearing on the
above conclusions.

\begin{figure*}
\hspace{-8mm}
  \includegraphics[width=184mm]{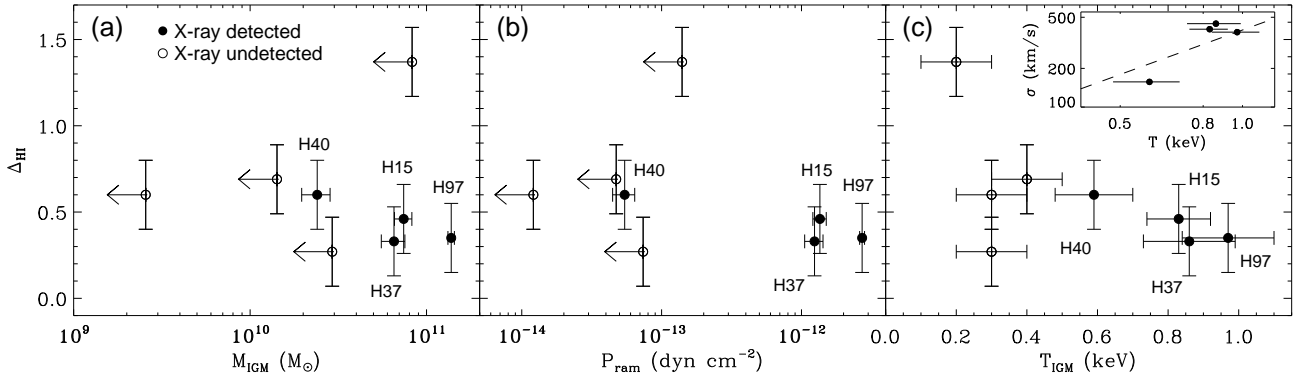}
  \caption{H{\sc i}~deficiency and (a) hot gas mass inside the region
  used for determining $\Delta_{\rm HI}$, (b) characteristic ram
  pressure, and (c) hot gas temperature for the various groups. Empty
  circles represent groups with no detectable hot gas. Inset in (c)
  shows velocity dispersion vs.\ $T_{\rm IGM}$ for the X-ray detected
  groups, with the \citet{osmo04} relation, i.e.\
  equation~(\ref{eq,sigma_T}), overplotted as a dashed line.}
\label{fig,hi_def}
\end{figure*}

Hence, neither is it surprising that $\Delta_{\rm HI}$ does not show a
clear dependence on the characteristic IGM ram pressure plotted in
Fig.~\ref{fig,hi_def}b and evaluated as the product of $\sigma_{\rm
V}^2$ and the volume-weighted mean IGM density $M_{\rm IGM}/V$ within
the volume $V=(4/3)\pi r_{\rm HI}^3$ covered by the radio data, with
all quantities taken from Table~\ref{tab,summary}. Note that the very
compact HCG\,40 -- in which we cannot completely rule out a residual
galactic contribution to the derived IGM mass -- stands out among the
X-ray detected groups, with a characteristic ram pressure 1--2 orders
of magnitude below that seen in the X-ray bright systems. For the
X-ray undetected groups, similar comments apply as for
Fig.~\ref{fig,hi_def}a.

Finally, in
Fig.~\ref{fig,hi_def}c we investigate the dependence of $\Delta_{\rm
HI}$ on IGM temperature. Thermal evaporation of galactic H{\sc i}
through heat conduction from the IGM is expected to proceed at a rate
$\dot{M} \propto T^{5/2}$ if unsuppressed by, for example, magnetic
fields. Given this strong temperature dependence, the lack of a
positive $\Delta_{\rm HI}$--$T$ correlation suggests that heat
conduction is not an important effect within our sample.  The location
of the exceptionally H{\sc i}~deficient HCG\,30 in
Fig.~\ref{fig,hi_def}c would seem to pose a particular challenge for
this mechanism.

Overall, Fig.~\ref{fig,hi_def} therefore seems to confirm the notion
that H{\sc i}~deficiency is not tightly linked to the presence or
nature of an IGM in these groups. There are some caveats to this
interpretation though. For example, it is worth emphasizing that it is
the four groups with the highest velocity dispersion that are X-ray
detected. If the remaining groups contain warm ($T\la 10^6$~K) rather
than hot gas, and thus fall well below the $\sigma$--$T$ relation for
X-ray bright systems, our constraints on the hot gas density could
seriously underestimate the true IGM density in these
systems. Unfortunately, this possibility cannot be directly tested
with the present data. However, the fact that our X-ray detected
systems scatter fairly tightly around the \citet{osmo04} relation, as
shown in the inset in Fig.~\ref{fig,hi_def}c, may support our use of
this relation for predicting $T$ also for the undetected
systems. Furthermore, results for other X-ray detected groups suggest,
if anything, that the poorest systems tend to have $T$ on the high
side for their velocity dispersion \citep{osmo04} although the
situation could, of course, be different for groups that remain X-ray
undetected.

Another concern is that $\sigma_{\rm V}$ and hence $P_{\rm ram}$ in
Fig.~\ref{fig,hi_def}b may not be robustly determined, being based on
just a handful of bright galaxies in most cases. Large corrections to
$\sigma_{\rm V}$ would be needed, however, in order to affect our
overall conclusions.  A further point is that any intergalactic H{\sc
i} already stripped from individual galaxies could potentially
contribute to the IGM mass and ram pressure. However, for the X-ray
detected systems, the {\em total} H{\sc i} mass inside the GBT beam is
of order 5~per~cent of the corresponding hot gas mass (Borthakur et
al., in prep.), suggesting that any cold gas can be neglected for the
present purposes.

Despite the appearance of Fig.~\ref{fig,hi_def}, it is premature to
exclude the possibility that galaxy--IGM interactions could play a
role for H{\sc i} removal in some of our groups.  For example, ram
pressure stripping is expected to occur when the IGM ram pressure
exceeds the gravitational restoring pressure of the galaxy. The
efficiency of this process therefore depends not only on the
properties of the IGM, but also on those of the individual group
galaxies. The process of viscous stripping \citep{nuls82} shares
similar features, and could be operating even when ram pressure itself
is insufficient to remove any H{\sc i}. Finally, ram pressure may also
indirectly affect the H{\sc i} in the disc. For example, many disc
galaxy formation models predict that massive spirals are surrounded by
hot gaseous haloes from which gas may cool out to provide fuel for
ongoing star formation in the disc (e.g., \citealt{toft02}). The
removal of this coronal gas by external forces could contribute to
H{\sc i}~deficiency, if the limited supply of H{\sc i} in the disc is
consumed by star formation without being replenished from the hot halo
(strangulation; see, e.g., \citealt{kawa08}). We next seek to quantify
the importance of these various mechanisms.

\section{The impact of the IGM: Modelling galaxy--IGM interactions}
\label{sec,model}

In an attempt to constrain the role of galaxy--IGM interactions for
typical disc galaxies in the individual groups, we constructed simple
analytical models of galaxies moving through the hot intragroup gas in
the gravitational potential of each group. As explained in
Section~\ref{sec,IGM}, the derived H{\sc i} deficiency for each group
should be viewed as an average for the group members, since it is
often non-trivial to evaluate observed H{\sc i} masses for the
individual members. For our modelling purposes, we have therefore
adopted a single, fiducial galaxy model, with overall properties
broadly matched to the fairly well-constrained mean properties of the
late-type group members in our sample. 

As described in detail below, the group potential and galaxy orbits
are less well determined for each group. Consequently, we evolve the
adopted galaxy model according to four different assumptions for each
group, corresponding to two choices for the gravitational potential,
and two for the galaxy orbits within the chosen potential.  The
variation among the resulting mass losses from the galaxy can serve as
a means to gauge the uncertainties associated with these
assumptions. We will consider two different stripping processes for
the cold gas in the disc, {\em viz.}\ classical ram pressure stripping
and turbulent viscous stripping. For the gas in any hot halo, we only
consider ram pressure stripping for simplicity.

\subsection{Group model and galaxy orbits}\label{sec,group}

Our goal is to evaluate the efficiency of ram pressure stripping and
related processes, without resorting to detailed numerical modelling,
which is beyond the scope of this work. For this purpose, we would
ideally adopt a single value of the ram pressure for each group.  The
simplest approach is to assume the classical analytical \citet{gunn72}
stripping criterion along with a constant ram pressure equal to its
peak value. Hydrodynamical simulations involving a constant ram
pressure have shown the \citet{gunn72} criterion to be remarkably
accurate in terms of predicting the disc stripping radius and the mass
of gas lost \citep*{abad99}, and it remains a reasonable approximation
even when allowing for orbital variations in ram pressure
\citep{jach07,roed07}.

However, recent hydrodynamical simulations of disc galaxies moving in
radial \citep{jach07} and two-dimensional orbits \citep{roed07} within
a non-uniform gas distribution have revealed an important exception to
this rule. If the ram pressure changes faster than the characteristic
stripping time-scale, as will be the case for galaxies moving through
a highly concentrated IGM, the \citet{gunn72} criterion tends to
overestimates the stripping efficiency. Since stripping is not
instantaneous, an ISM element may not always be accelerated to
galactic escape velocity before the peak of the ram pressure is
over. In such cases the gas will eventually re-accrete, a possibility
not taken into account by the \citet{gunn72} criterion. In the present
study, these considerations could potentially be relevant for several
of our groups, given the fairly small core radii resulting from the
surface brightness fits. Instead of using the peak value of the ram
pressure in our modelling, it therefore seems more sensible to adopt
an orbit-averaged mean ram pressure by evaluating the time spent by
the galaxy at a given velocity and IGM density.

To this end, we first derived total mass profiles $M_{\rm tot}(r)$ for
each X-ray detected group, assuming a spherically symmetric gas
distribution in hydrostatic equilibrium,
\begin{equation}
 M_{\rm tot}(<r) = \frac{-kT(r)r}{G\mu m_p} \left(
 \frac{\mbox{d\,ln\,} n(r)}{\mbox{d\,ln\,} r} + \frac{\mbox{d\,ln\,}
 T(r)}{\mbox{d\,ln\,} r} \right).
\label{eq,mass}
\end{equation}
Since in general we have neither the statistics nor the spatial
coverage to constrain $T(r)$ to large radii, we made two assumptions
about the temperature profile which are likely to bracket the actual
temperature distribution in these somewhat disturbed groups. We either
simply assumed $T(r)$ equal to a constant mean value $\langle T
\rangle$, or $T(r)= -0.5\langle T\rangle \mbox{ log}(r/r_{500}) +
0.67$, appropriate for reasonably undisturbed groups outside any cool
core \citep{rasm07}. In the latter case, $r_{500}$, the radius
enclosing a mean density of 500 times the critical value $\rho_c$, was
evaluated iteratively until convergence was reached. For $\langle
T\rangle$ we used the measured value listed in
Table~\ref{tab,summary}.

The resulting mass profiles were characterized analytically by fitting
them with standard `NFW' models \citep*{nava97}, in which the dark
matter distribution is described by
\begin{equation}
	\rho(r) = \rho_c \frac{\delta_c}{(r/r_s)(1+r/r_s)^2},
\label{eq,NFW}
\end{equation}
where $\delta_c$ is a dimensionless parameter related to total group
mass $M_{200}$, and $r_s$ is a scale radius which reflects the more
commonly used concentration parameter $c=r_{200}/r_s$ (see
\citealt{nava97} for details). With $M_{200}$, and hence $\delta_c$,
fixed from the measured (extrapolated) mass profile,
equation~(\ref{eq,NFW}) was fitted to this profile to derive values of
$c$ under the two assumptions about $T(r)$ described above. The
results, summarized in Table~\ref{tab,nfw}, imply typical values of
$M_{200}\approx 1-2\times 10^{13}$~M$_\odot$ with $c\sim 5-10$.  For
these group masses, derived concentration parameters are thus in good
agreement with expectations from cosmological $N$-body simulations
\citep{bull01}. Note that the assumption of a declining temperature
profile typically reduces the derived group mass by $\sim 30$~per~cent
while increasing the halo concentration by a factor of $\sim 2$.

\begin{table} 
 \centering 
   \caption{Results of NFW fits to the derived group mass profiles
    under the two assumptions for $T(r)$ described in the text.}
  \label{tab,nfw} 
\begin{tabular}{@{}lcccr@{}} 
  \multicolumn{5}{c}{{\em T isothermal}} \\ \hline
  \multicolumn{1}{l}{Group} &
  \multicolumn{1}{c}{$r_{500}$} &
  \multicolumn{1}{c}{$r_{200}$} &
  \multicolumn{1}{c}{$M_{200}$} &
  \multicolumn{1}{c}{$c$}  \\
  & (kpc) & (kpc) & ($10^{13}$~M$_\odot$) &  \\ \hline
  HCG\,15 & 342 & 541 & 2.0 & 4.5 \\
  HCG\,37 & 362 & 572 & 2.3 & 4.8 \\
  HCG\,40 & 357 & 565 & 2.2 & 4.0 \\
  HCG\,97 & 379 & 500 & 2.7 & 4.7 \\ \hline
\end{tabular}

\vspace{2mm}

\begin{tabular}{@{}lcccr@{}} 
  \multicolumn{5}{c}{{\em T declining}} \\ \hline
  \multicolumn{1}{l}{Group} &
  \multicolumn{1}{c}{$r_{500}$} &
  \multicolumn{1}{c}{$r_{200}$} &
  \multicolumn{1}{c}{$M_{200}$} &
  \multicolumn{1}{c}{$c$}  \\
  & (kpc) & (kpc) & ($10^{13}$~M$_\odot$) &  \\ \hline
  HCG\,15 & 317 & 474 & 1.3 & 10.0 \\
  HCG\,37 & 334 & 499 & 1.6 & 11.0 \\
  HCG\,40 & 318 & 474 & 1.3 &  9.0 \\
  HCG\,97 & 349 & 523 & 1.8 & 10.9 \\ \hline
\end{tabular}

\end{table}

For the galaxy orbital configuration within the derived gravitational
NFW potential, we assume two different orbits, both radial and hence
going through the group core. The galaxy is assumed to be experiencing
a face-on IGM encounter in either case. These maximizing assumptions
allow us to estimate how important ram pressure stripping can ideally
be in our groups. The two orbits differ only in terms of the assumed
initial position and velocity of the galaxy, with the galaxy initially
at rest at a small clustercentric radius in the first scenario, and
falling towards the group core from a large radius and with a high
initial velocity in the second. Specifically, the following two
scenarios are considered:
\begin{itemize}
\item[(i)] For each group, we determine $\bar r$, the observed
(projected) mean clustercentric distance of the principal galaxies in
each group.  The model galaxy is assumed to be initially at rest at a
larger clustercentric distance $r_0$, from which it falls freely
towards the group centre. $r_0$ is chosen such that when the galaxy
reaches $r = \bar r$, it has attained a velocity corresponding to the
observed group velocity dispersion. Typical values are  $r_0 \sim 100$~kpc and
$\bar r\sim 40$~kpc.
\item[(ii)] The galaxy enters the group halo at $r_0=r_{200}$ with a
radial velocity $v_r$ corresponding to the halo circular velocity at
this radius, $v_r = (GM/r_{200})^{1/2}$. Typical values are $r_0\sim 500$~kpc
and $v_r\sim 400$~km~s$^{-1}$.
\end{itemize}
In both cases we follow the galaxy until it turns around, having
completed one passage through the group core. The first scenario is
chosen in an effort to match the observed average position (modulo
projection effects) and galaxy velocity in each group at present. In
practice, it represents galaxy motion fairly close to the group core,
with a mildly varying ram pressure, and so is somewhat reminiscent of
the `classical' ram pressure scenario involving a constant, high ram
pressure.  The assumption underlying this orbit is extreme, however,
in the sense that the true clustercentric distances will generally be
larger than the observed (projected) ones, which implies that galaxies
will generally spend a larger fraction of their time at large distance
than implied by this assumption. Therefore we also consider scenario
(ii) as a kind of opposite extreme. In this, galaxies experience a
considerably higher peak ram pressure, but this occurs only relatively
briefly. In the following, these two scenarios will be referred to as
orbit (i) and (ii), respectively.

Although these two orbits do not necessarily encompass the extreme
orbital solutions for the group members, they do represent two rather
different cases, thus offering a handle on the uncertainty in the
predicted mass loss related to orbital assumptions. Furthermore, while
completely radial orbits may not be very common, we note that H{\sc i}
deficient spirals in clusters tend to have more eccentric orbits than
non-deficient ones \citep{sola01}, and that cosmological infall along
filaments would proceed in fairly eccentric orbits, thus lending some
support to our simplifying assumption. For a detailed discussion of
the impact of orbital parameters on the stripping efficiency, we refer
to \citet{hest06}.

\subsection{Galaxy model}

For the galaxy model, needed to estimate the gravitational restoring
force and hot halo thermal pressure of a `typical' late-type group
galaxy within our sample, we follow the general approach described in
\citet{rasm06} which is repeated here for completeness. The model
consists of a spherical dark matter (DM) halo with density profile
\begin{equation} 
  \rho_{\rm h}(r)=\frac{M_{\rm h}}{2\pi^{3/2}} \frac{\eta}{r_{\rm t}  
    r_{\rm h}^2} \frac{\mbox{exp}(-r^2/r_{\rm t}^2)}{(1+r^2/r_{\rm h}^2)}, 
\label{eq,DM}
\end{equation} 
a hot gaseous halo of the same form, a spherical bulge with
\begin{equation}
  \rho_{\rm b}(r)=\frac{M_{\rm b}}{2\pi r_{\rm b}^2}
  \frac{1}{r(1+r/r_{\rm b})^3},
\end{equation}
and exponential stellar and gaseous discs, each of the form 
\begin{equation} 
  \rho_{\rm d}(R,z) = \frac{M_{\rm d}}{4\pi R_{\rm d}^2 z_{\rm d}}  
  \mbox{exp}(-R/R_{\rm d})\mbox{sech}^2(z/z_{\rm d}). 
\end{equation}  
Here $M_{\rm b}$, $M_{\rm h}$, and $M_{\rm d}$ are the total masses of
each component, $r_{\rm b}$ and $r_{\rm h}$ are the scalelengths of
the bulge and halo, respectively, $r_{\rm t}$ is the DM halo
`truncation' radius, $R_{\rm d}$ is the cylindrical scalelength of the
disc components and $z_{\rm d}$ the corresponding thickness, and
finally
\begin{equation} 
\eta = \{{1 - \pi^{1/2}q\mbox{\,exp}(q^2)[1-\mbox{erf}(q)]}\}^{-1}, 
\end{equation} 
where $q=r_{\rm h}/r_{\rm t}$ and erf is the error function. With this
model, the restoring gravitational acceleration $\frac{\partial
\Phi}{\partial z}(R,z)$ in the direction $z$ perpendicular to the disc
can be evaluated analytically for each model component using the
equations of \citet{abad99}, to whom we refer for more details.

In order to constrain model parameters, stellar masses of the HCG
members were evaluated from their $B$- and $K$-band magnitudes as
listed in NED, following the prescription adopted by \citet{mann05}.
For this purpose, only the principal members in our eight groups, as
listed by \citet{hick82}, with morphological types later than S0 were
included, yielding a mean stellar mass of $4.1\times
10^{10}$~M$_\odot$. For a subset of these galaxies (8 out of 27),
maximum disc rotational velocities, indicative of total galaxy masses,
are also available in the Hyperleda database, with a mean value of
135~km~s$^{-1}$. We note that, in terms of mean stellar mass, this
subset is representative of the full sample, showing $\langle
M_{\ast}\rangle=4.4\times 10^{10}$~M$_\odot$. We therefore assume a
stellar mass of $4\times 10^{10}$~M$_\odot$ for the galaxy model,
distributed such as to yield a bulge-to-disc mass ratio of 1/4,
appropriate for an Sb/c spiral. Using the relation of \citet{hayn84},
we further assume an H{\sc i} mass of $6.7\times 10^9$~M$_\odot$ to
ensure that our model galaxy initially has a `normal' H{\sc i} mass
for its sample-averaged blue luminosity of $L_B=1.6\times
10^{10}$~L$_\odot$.

The scalelengths of the stellar and gaseous disc components in the
model were chosen to ensure that at least 85~per~cent of the stellar
and H{\sc i} mass resides within the average optical disc radius of
$r_D \simeq 10$~kpc, as derived from the size of the $D_{25}$ ellipse
for each spiral member in our groups. For simplicity, the gas
distribution in the hot gaseous halo is assumed to follow that of the
underlying dark matter, but with a smaller value of $r_t$
corresponding to twice the `size' $r_D$ of the stellar disc, and with
a total mass corresponding to the H{\sc i} mass in the disc. The hot
halo is assumed to be isothermal at a temperature $T\sim 0.06$~keV,
the virial temperature corresponding to the maximum allowed disc
rotational velocity in the model, taken to be $\sim 150$~km~s$^{-1}$.
Once the baryonic model components have been specified, the parameters
of the DM halo are effectively set by this maximum allowed rotation
velocity, and by the requirement that the model baryon fraction match
the universal value of $\sim 15$~per~cent.  Table~\ref{tab,model}
summarises the adopted model galaxy parameters.  Note that the cold
gas component in the model refers to the distribution of H{\sc i}
only, as we do not consider any molecular gas here. In summary, this
model roughly reproduces the average stellar mass, bulge-to-disc
ratio, disc size, and maximum disc rotational velocity seen for the
spirals in our groups, with a total baryon fraction consistent with
the universal value, and an initial H{\sc i} mass as expected for a
non-stripped isolated galaxy with these properties.

\begin{table}
\begin{center}
  \caption{Adopted parameters of the galaxy model. $L$ is the
    characteristic scalelength for each component, i.e.\ $r_{\rm h}$
    and $r_{\rm t}$ for the haloes, $r_{\rm b}$ for the bulge, and
    $R_{\rm d}$ and $z_{\rm d}$ for the disc components.}
\label{tab,model}
 \begin{tabular}{@{}lcl}
   \hline
   Component     & $M_{\rm total}$ & $L$        \\ 
                 & ($10^{10}$~M$_\odot$) & (kpc) \\ \hline
   DM halo       &  28        & 5, 100  \\
   Hot gas halo  &   0.7      & 5, 20 \\
   Stellar bulge &   0.8      & 0.5     \\
   Stellar disc  &   3.2      & 3, 0.25 \\         
   Cold gas disc &   0.67     & 3, 0.25 \\
   \hline
\end{tabular}
\end{center}
\end{table}

\subsection{Computing mass losses}\label{sec,massloss}

Combining equation~(\ref{eq,NFW}) with the measured density
distribution of intragroup gas, time--averaged values of IGM density
$\langle \rho \rangle$ and the square of the orbital velocity $\langle
v_{\rm gal}^2 \rangle$ experienced by the galaxy in its orbit can be
evaluated. For the averaging time-scale, we only consider the segment
of the orbit for which the ram pressure exerts a significant influence
on the ISM, taken to be from the time at which the instantaneous ram
pressure, if sustained, would remove at least 1~per~cent of the cold
gas. This is to avoid the artificial suppression of $\langle \rho
\rangle$ and $\langle v^2 \rangle$ that would otherwise result from
including the time spent by the galaxy at large radius where ram
pressure is completely negligible.  Note that the corresponding
characteristic ram pressure is independent of group mass and orbital
initial conditions, as it depends only on the assumed galaxy model.

Assuming a face-on IGM encounter, the condition for ram-pressure
stripping is then evaluated as
\begin{equation}
 \Sigma_{\rm g}\left(\frac{\partial \Phi_{\rm b}}{\partial z} + 
 \frac{\partial \Phi_{\rm h}}{\partial z} +
 \frac{\partial \Phi_{\rm g}}{\partial z} +
 \frac{\partial \Phi_{\ast}}{\partial z} \right)
 < \langle \rho \rangle \langle v_{\rm gal}^2\rangle ,
\label{eq,strip}
\end{equation}
where $\frac{\partial \Phi}{\partial z}(R,z)$ is the restoring
gravitational acceleration in the direction $z$ perpendicular to the
disc, originating from the stellar bulge (subscript 'b'), dark matter
halo ('h'), gaseous disc ('g') and stellar disc ('$\ast$'),
respectively, and $\Sigma_{\rm g}$ is the surface density of cold gas.
Equation~(\ref{eq,strip}) is similar to the classical \citet{gunn72}
stripping criterion, but takes into account the mass distribution in
the galaxy rather than simply assuming a homogeneous disc thin enough
to be described solely by its surface density. Its solution also
provides us with the `stripping region', the region in the ($R$,
$z$)--plane from which gas is permanently lost by the galaxy.

Apart from conventional ram pressure stripping, transport processes
such as viscous stripping \citep{nuls82}, caused by Kelvin--Helmholtz
instabilities arising at the ISM--IGM interface, could also play a
role even when the ram pressure itself is insufficient to remove
galactic gas (\citealt*{quil00}; \citealt{rasm06}). Turbulent viscous
stripping of a gas disc of radius $r_D$ is expected to operate at
Reynolds numbers $Re \ga 30$, where
\begin{equation}
Re = \mbox{$\cal M$} r_D/\lambda,
\label{eq,Re}
\end{equation}
$\cal M$ is the Mach number of the IGM flow past the galaxy, and
$\lambda \propto T^2 n^{-1}$ is the ion mean free path in the IGM.
The expected mass-loss rate due to this process \citep{nuls82},
\begin{equation}
\dot M_{\rm vs} \approx 0.5 \pi r_D^2 \rho v_{\rm gal}, 
\label{eq,vs}
\end{equation}
scales only linearly with galaxy velocity and so could be important in
a wider range of environments than ram pressure itself. We include
this process in the stripping calculations by evaluating
equations~(\ref{eq,Re}) and (\ref{eq,vs}) at each point in the orbit
and adding up the total mass loss. Dealing with disc galaxies rather
than spheroids, we have used only half the \citet{nuls82} mass loss
rate in equation~(\ref{eq,vs}) because of the correspondingly smaller
galaxy surface area for a given $r=r_D$. For $r_D$ itself, we use the
smaller of the sample-averaged value of $r_D=10$~kpc and the stripping
radius predicted by equation~(\ref{eq,strip}).  Note that ${\cal M}
\le 1$ in equation~(\ref{eq,Re}) even if the galaxy is moving
supersonically, as the post-shock IGM flow past the galaxy will always
be subsonic. Also note that the stripping efficiency of both ram
pressure and viscous stripping should be largely unaffected by the
presence of a shock front (see \citealt{rasm06}).

Having dealt with the stripping of cold gas from the disc, we now turn
to the removal of any gas situated in a hot galactic halo.
It seems plausible that, at the very
least, the cooling and gradual inflow of any such gas will be disrupted
once the external ram pressure $P_{\rm ram}$ exceeds its thermal
pressure $P_{\rm th}$. The simulations of \citet{mori00} confirm that
this condition provides a reasonable estimate of the mass of a galaxy
that will have its hot halo completely stripped by ram pressure. For the
purpose of also assessing the importance of strangulation for the
model galaxy, we therefore use the derived values of $\langle \rho
\rangle$ and $\langle v^2 \rangle$ to evaluate the `strangulation
region' where $P_{\rm ram} > P_{\rm th}$, and the corresponding mass
of affected coronal gas. The aim here is only to develop a rough
picture of the potential importance of IGM interactions for
strangulation, so the effects of viscous stripping are not considered
for the hot halo.

We readily acknowledge that our modelling approach is inferior to
detailed numerical models (e.g., \citealt{hest06}), and hydrodynamical
simulations in particular. It is our hope that one can nevertheless
have some confidence in the results, given that we have employed a
reasonably detailed galaxy model and are making some allowance for
orbital variations in ram pressure. The advantage of the adopted
method is that it is easily tailored to the specific conditions in
individual groups at little computational cost.

\subsection{Results}

We re-iterate that we are considering four scenarios for each group,
consisting of two separate assumptions about the galaxy orbit, and two
about the IGM temperature (and hence total mass) distribution in the
groups, as specified in Section~\ref{sec,group}. The results of the
H{\sc i} stripping calculations for each of these four cases are
summarized in Table~\ref{tab,strip}, which lists the derived r.m.s.\
orbital velocity $v_{\rm rms}$, the orbit-averaged IGM number density,
the mass of H{\sc i} lost due to ram pressure and viscous stripping,
and the total fraction of H{\sc i} stripped from the initial model
reservoir of $6.7\times 10^9$~M$_\odot$. Recall that $v_{\rm rms}$ is
computed only from the point in the orbit where ram pressure becomes
significant and so depends not only on orbital parameters but also on
the IGM distribution.

\begin{table} 
 \centering 
   \caption{Orbit--averaged galaxy velocities and IGM densities, along
   with predicted H{\sc i} mass loss $\Delta M$ due to ram pressure
   (`rp') and viscous stripping (`vs') after one passage through the
   group core for the assumed orbits and group mass profiles. The
   final column lists the total fraction $f$ of H{\sc i} lost.}
  \label{tab,strip} 
\begin{tabular}{@{}lccccc@{}} 
  \multicolumn{6}{c}{{\em Orbit (i), T isothermal}} \\ \hline
  \multicolumn{1}{l}{Group} &
  \multicolumn{1}{c}{$v_{\rm rms}$} &
  \multicolumn{1}{c}{$\langle n\rangle$} &
  \multicolumn{1}{c}{$\Delta M_{\rm rp}$} &
  \multicolumn{1}{c}{$\Delta M_{\rm vs}$} & 
  \multicolumn{1}{c}{$f$} \\
  & (km~s$^{-1}$) & (cm$^{-3}$) & ($10^9$~M$_\odot$) & ($10^9$~M$_\odot$) 
  &  \\ \hline
  HCG\,15 & 354 & $6.7\times 10^{-4}$ & 0.9 & 1.0 & 0.29 \\
  HCG\,37 & 342 & $1.5\times 10^{-3}$ & 1.2 & 1.4 & 0.39 \\
  HCG\,40 & 199 & $7.8\times 10^{-4}$ & 0.6 & 0.2 & 0.12 \\
  HCG\,97 & 362 & $2.6\times 10^{-3}$ & 1.6 & 2.3 & 0.58 \\ \hline
\end{tabular}

\vspace{2mm}

\begin{tabular}{@{}lccccc@{}} 
  \multicolumn{6}{c}{{\em Orbit (i), T declining}} \\ \hline
  \multicolumn{1}{l}{Group} &
  \multicolumn{1}{c}{$v_{\rm rms}$} &
  \multicolumn{1}{c}{$\langle n\rangle$} &
  \multicolumn{1}{c}{$\Delta M_{\rm rp}$} &
  \multicolumn{1}{c}{$\Delta M_{\rm vs}$} & 
  \multicolumn{1}{c}{$f$} \\
  & (km~s$^{-1}$) & (cm$^{-3}$) & ($10^9$~M$_\odot$) & ($10^9$~M$_\odot$) 
  & \\ \hline
  HCG\,15 & 404 & $7.2\times 10^{-4}$ & 1.0 & 0.9 & 0.29 \\
  HCG\,37 & 402 & $1.9\times 10^{-3}$ & 1.4 & 1.2 & 0.39 \\
  HCG\,40 & 266 & $8.2\times 10^{-4}$ & 0.8 & 0.2 & 0.14 \\
  HCG\,97 & 435 & $2.8\times 10^{-3}$ & 1.8 & 2.0 & 0.57 \\ \hline
\end{tabular}

\vspace{2mm}

\begin{tabular}{@{}lccccc@{}} 
  \multicolumn{6}{c}{{\em Orbit (ii), T isothermal}} \\ \hline
  \multicolumn{1}{l}{Group} &
  \multicolumn{1}{c}{$v_{\rm rms}$} &
  \multicolumn{1}{c}{$\langle n\rangle$} &
  \multicolumn{1}{c}{$\Delta M_{\rm rp}$} &
  \multicolumn{1}{c}{$\Delta M_{\rm vs}$} & 
  \multicolumn{1}{c}{$f$} \\
  & (km~s$^{-1}$) & (cm$^{-3}$) & ($10^9$~M$_\odot$) & ($10^9$~M$_\odot$) 
  &  \\ \hline
  HCG\,15 & 498 & $1.2\times 10^{-4}$ & 0.6 & 1.2 & 0.27 \\
  HCG\,37 & 541 & $1.5\times 10^{-4}$ & 0.7 & 1.9 & 0.39 \\
  HCG\,40 & 630 & $4.3\times 10^{-5}$ & 0.5 & 0.3 & 0.12 \\
  HCG\,97 & 536 & $2.8\times 10^{-4}$ & 0.9 & 4.3 & 0.77 \\ \hline
\end{tabular}

\vspace{2mm}

\begin{tabular}{@{}lccccc@{}} 
  \multicolumn{6}{c}{{\em Orbit (ii), T declining}} \\ \hline
  \multicolumn{1}{l}{Group} &
  \multicolumn{1}{c}{$v_{\rm rms}$} &
  \multicolumn{1}{c}{$\langle n\rangle$} &
  \multicolumn{1}{c}{$\Delta M_{\rm rp}$} &
  \multicolumn{1}{c}{$\Delta M_{\rm vs}$} & 
  \multicolumn{1}{c}{$f$} \\
  & (km~s$^{-1}$) & (cm$^{-3}$) & ($10^9$~M$_\odot$) & ($10^9$~M$_\odot$) 
  &  \\ \hline
  HCG\,15 & 448 & $1.2\times 10^{-4}$ & 0.5 & 1.2 & 0.26 \\
  HCG\,37 & 491 & $1.5\times 10^{-4}$ & 0.6 & 1.9 & 0.38 \\
  HCG\,40 & 554 & $4.6\times 10^{-5}$ & 0.4 & 0.3 & 0.11 \\
  HCG\,97 & 477 & $2.8\times 10^{-4}$ & 0.8 & 4.3 & 0.76 \\ \hline
\end{tabular}

\end{table} 

As can be seen from the Table, the amount of gas lost through either
stripping process is generally an appreciable fraction of the initial
H{\sc i} mass. Viscous stripping in particular can remove a
substantial fraction of the cold disc gas for all model
assumptions. HCG\,40 is the exception, with the relatively tenuous IGM
in this group removing at most 10--15~per~cent of the H{\sc i}. The
steeper IGM density profile in this group also implies that a larger
fraction of the total IGM mass is encountered at high velocity,
leading to a higher $\Delta M_{\rm rp}/\Delta M_{\rm vs}$ ratio than
for the other groups, and a higher value of $v_{\rm rms}$ in orbit
(ii), because the ram pressure becomes significant closer to the core
in this system.

The Table further shows that the main variation in the computed mass
loss for a given group and stripping mechanism derives from the choice
of orbit rather than the assumed group mass profile. Orbit (i) is
generally more efficient at removing gas through ram pressure
stripping than orbit (ii), because although the peak ram pressure is
considerably higher in the latter case, the galaxy spends a
comparatively shorter time in regions corresponding to high values of
$P_{\rm ram}$. Conversely, viscous stripping is more efficient in
orbit (ii), as this mechanism acts even at relatively low $v_{\rm
gal}$, allowing the associated mass loss to build up significantly
over the much longer crossing time-scale relevant for this orbit.
Note, however, that for a given group, the outcome in terms of {\em
total} H{\sc i} mass loss, $\Delta M_{\rm tot} = \Delta M_{\rm rp} +
\Delta M_{\rm vs}$, is largely insensitive to the various orbit and
mass profile assumptions, despite clear variations in $\langle n
\rangle$ and $v_{\rm rms}$ among the four scenarios. The only
exception to this is the significantly more massive HCG\,97, for which
the deeper gravitational potential and higher IGM mass implies that
orbit (ii) is relatively more efficient at removing galactic gas than
for the other groups.

Figure~\ref{fig,strip} outlines the stripping region for a model
galaxy in each of the X-ray detected groups, based on solving
equation~(\ref{eq,strip}). The figure shows the one of our four
scenarios in which the effect of ram pressure is generally most
pronounced, i.e., orbit (i) with $T(r)$ declining. In the outer disc,
the gravitational restoring pressure, and hence the stripping region
for the cold gas, is seen to be nearly independent of vertical disc
height for interesting values of $|z|$, and the gravitational
restoring force at a given $R$ peaks well above the disc. Compared to
the `size' of the stellar disc ($r_D=10$~kpc), it is clear that the
gas disc becomes mildly truncated by ram pressure in HCG\,97 and 37
but remains largely unaffected in the other two groups. Note that this
truncation reduces the viscous stripping efficiency by reducing the
surface area of the gas disc exposed to the IGM.  Viscous stripping is
therefore slightly more efficient for low values of $P_{\rm ram}$,
adding to the explanation of the higher values of $\Delta M_{\rm vs}$
for orbit (ii) in Table~\ref{tab,strip}.

\begin{figure}
  \includegraphics[width=84mm]{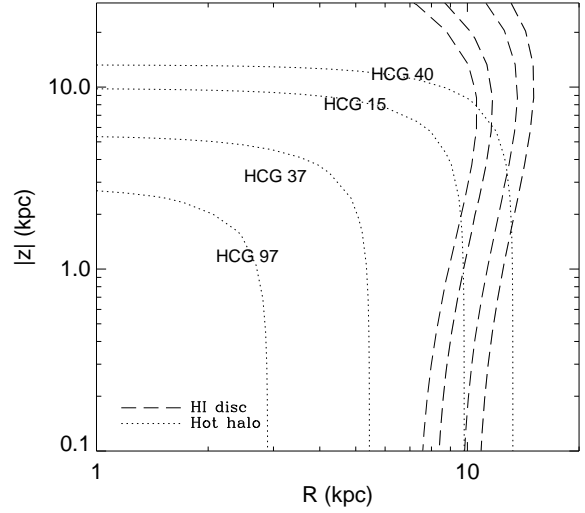}
  \caption{Isobars of gravitational restoring pressure (dashed) and
   hot halo thermal pressure (dotted) for our fiducial model galaxy in
   the different groups. Dashed lines outline the galactic regions
   outside which H{\sc i} in the disc can be stripped by ram pressure,
   corresponding to equality in equation~(\ref{eq,strip}). Dotted
   lines show the corresponding regions for the hot halo gas. The
   order of the contours is the same in both cases.}
\label{fig,strip}
\end{figure}

Predicted H{\sc i}~deficiencies corresponding to the mass losses in
Table~\ref{tab,strip} are compared to the observed values in
Fig.~\ref{fig,massloss}. Errors on predicted deficiencies correspond
to the full range in predicted H{\sc i} mass loss for each group under
the various orbit and mass profile assumptions.  Shown are the
expectations from ram pressure stripping alone, as well as the full
H{\sc i} mass loss from combining equations~(\ref{eq,strip}) and
(\ref{eq,vs}). The dashed line in the figure represents equality
between modelled and observed H{\sc i}~deficiencies; under our model
assumptions, anything below this line cannot be explained by
galaxy--IGM interactions alone.  The plot clearly suggests that ram
pressure stripping on its own is not sufficient to cause the observed
H{\sc i}~deficiencies, even in this X-ray bright subsample of our
groups. When including viscous stripping, however, if as efficient in
removing H{\sc i} as assumed here, galaxy--IGM interactions can
certainly help explain observed values of $\Delta_{\rm HI}$, but they
can potentially fully account for the H{\sc i} loss only in HCG\,37
and HCG\,97, i.e.\ in just two out of our eight groups.

\begin{figure}
 \hspace{0mm}
  \includegraphics[width=82mm]{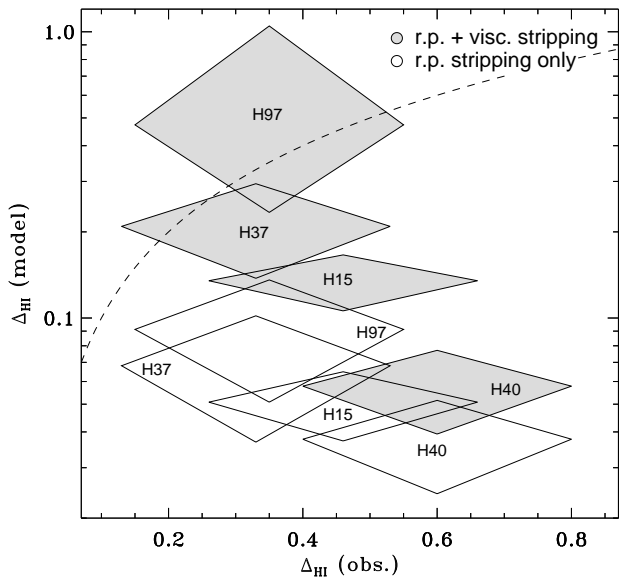}
  \caption{Observed H{\sc i}~deficiencies compared to the model
  predictions of Table~\ref{tab,strip} for the stripping of H{\sc i}
  by ram pressure alone (empty diamonds), and by ram pressure plus
  viscous stripping (shaded). Dashed line represents equality between
  observed and predicted $\Delta_{{\rm HI}}$.}
\label{fig,massloss}
\end{figure}

Regarding the issue of strangulation, Fig.~\ref{fig,strip} also
illustrates the derived stripping region for hot halo gas according to
the adopted $P_{\rm ram} > P_{\rm th}$ criterion. The figure suggests
that ram pressure alone could remove a sizable fraction of the halo
gas in the model. Table~\ref{tab,corona} lists the fractions of
stripped halo gas in the different scenarios, showing that these can
be substantial for orbit (i) in particular, peaking at $\sim
95$~per~cent in HCG\,97. Viscous stripping, not included here, could
potentially contribute beyond these estimates. This suggests that
galaxy--IGM interactions in X-ray bright groups could play an
important role in removing the gas supply that may ultimately fuel
star formation in spirals, in qualitative agreement with the
simulation results of \citet{kawa08}. For our specific model setup,
there is a large dispersion in the fraction of gas affected, however,
and it is not clear that the effect would be important in groups such
as HCG\,40.

\begin{table} 
 \centering 
   \caption{Fraction of hot halo gas lost due to ram pressure in the
            various orbital scenarios.}
  \label{tab,corona} 
\begin{tabular}{@{}lcccc@{}} \hline
  \multicolumn{1}{l}{Group} &
  \multicolumn{1}{c}{Orbit (i)} &
  \multicolumn{1}{c}{Orbit (i)} &
  \multicolumn{1}{c}{Orbit (ii)} &
  \multicolumn{1}{c}{Orbit (ii)}  \\
  & $T(r)$ const. & $T(r)$ decl.  & $T(r)$ const & $T(r)$ decl. \\ \hline
  HCG\,15 & 0.53 & 0.64 & 0.31 & 0.27 \\
  HCG\,37 & 0.74 & 0.85 & 0.39 & 0.35 \\
  HCG\,40 & 0.31 & 0.46 & 0.21 & 0.18 \\
  HCG\,97 & 0.84 & 0.94 & 0.52 & 0.48 \\ \hline
\end{tabular}
\end{table}

\subsection{Model limitations and caveats}\label{sec,caveats}

It is important to stress that the calculations presented here are
only intended to provide a rough picture of the impact of galaxy--IGM
interactions in our sample, and we do not claim that these results are
anything but indicative. Caveats include the fact that the adopted
galaxy model parameters provide a plausible, but not necessarily
unique, representation of the spirals in our sample.  It is also a
possibility, albeit one we cannot easily evaluate, that some group
members may be individually less H{\sc i} deficient than the
`group-averaged' value of $\Delta_{\rm HI}$, and so could have H{\sc
i} deficiencies consistent with removal by ICM interactions alone,
even if our model results suggest otherwise for the group members as a
whole.

As regards the model calculations, an important limitation is the fact
that the model is completely static. In practice, the gas
distributions both in the groups and their galaxies will be evolving,
which could be particularly relevant for orbit (ii) where orbital
time-scales are several Gyr. Also, the estimate of viscous stripping
mass loss does not take into account that this process itself will
reduce the size of the gas disc and hence the mass loss rate according
to equation~(\ref{eq,vs}). Furthermore, this stripping process may
saturate \citep{nuls82} which would reduce the mass loss, the presence
of magnetic fields could also suppress hydrodynamical instabilities,
and the presence of a hot halo could perhaps to some extent shield the
cold gas disc from such instabilities. The latter could be of
particular relevance for orbit (ii), in which $\Delta M_{\rm vs}$ is
high but where a comparatively smaller fraction of halo gas is lost
due to ram pressure. These considerations suggest that the estimated
contribution by viscous stripping to the H{\sc i} mass loss should be
regarded as an upper limit.

With their compact galaxy configurations and low velocity dispersions,
our groups could also represent environments in which galactic dark
matter haloes are subject to significant tidal truncation. Hence,
another issue is to what extent our results are affected by the
adopted assumptions on dark matter in the galaxy model. For the
adopted disc and bulge parameters, our freedom to modify the assumed
DM distribution is mainly constrained by the requirement that the
maximum disc rotational velocity $v_{\rm max}$ of the model should not
exceed the allowed $\sim 150$~km$^{-1}$.  A full exploration of model
parameter space is beyond the scope of this work, and we refer to
\citet{hest06} for a more thorough investigation of these
issues. Nevertheless, to provide a rough picture of the impact of DM
for our results, we repeated the stripping calculations of
Section~\ref{sec,massloss} with three different modifications, subject
to the above $v_{\rm max}$ criterion:
\begin{itemize}
\item[(i)] Assuming no DM [or larger DM halo scalelengths $r_t$ and
$r_h$ in equation~(\ref{eq,DM})]. This must be regarded as an extreme
assumption, which should facilitate stripping.
\item[(ii)] Increasing the DM mass by a factor of two (for which the
$v_{\rm max}$ criterion then
requires at least one of the DM scalelengths $r_t$ and $r_h$ to
increase by a similar factor). This should suppress stripping.
\item[(iii)] Assuming a higher DM concentration (i.e.\ lower $r_h$ or
$r_t$) by a factor of three (suppresses stripping), which in turn
requires a lower total DM mass by a factor of five (facilitates
stripping). These choices comply with the typical tidal truncation of
DM halos inferred for galaxies in massive clusters \citep{limo07}.
\end{itemize}
We find that, in all three cases, the amount of H{\sc i} lost by
the model galaxy is at most modified by 10--15~per~cent, regardless of
the group and orbit considered. This perhaps somewhat surprising
result has its origin in the fact that the vast majority of cold gas
in the model is located at low galactocentric distances, where the
restoring gravitational acceleration -- which determines the stripping
region in accordance with equation~(\ref{eq,strip}) -- is dominated by
the baryonic disc components and not the dark matter halo.  The
variations in the stripped gas mass for the hot halo can be larger, up
to 45~per~cent for HCG\,40, but they can generally easily be
accommodated by the variations associated with the different orbital
assumptions. We therefore conclude that our results are reasonably
robust to changes in the assumed amount and distribution of dark
matter in the galaxy model.

\section{Discussion}\label{sec,discus}

The groups in our sample are all H{\sc i}~deficient even when
accounting for any intergalactic H{\sc i} not clearly associated with
individual galaxies. Removal of H{\sc i} from the group members is by
itself insufficient to explain this situation, as the removed gas must
also be prevented from staying neutral. Thus, unless H{\sc i} is
somehow destroyed {\em in situ} within the galaxies, it must go
through a two-stage process whereby it is first removed from the
galaxies and then ionized by a possibly unrelated mechanism. Here we
discuss these different possibilities in light of our X-ray and
modelling results.

\subsection{{\em In situ} destruction of H{\sc i}}

The H{\sc i} could potentially have been destroyed within the galaxies
themselves, either by evaporation though thermal conduction from the
IGM, or through consumption by star formation, provided there is no
continuous replenishment of cold material. A third possible mechanism
involves heating and possibly ejection from the galaxies by starburst
winds. While the first possibility, direct heating by the IGM, seems
implausible in light of the absence of a positive correlation between
$\Delta_{\rm HI}$ and $T_{\rm IGM}$ (cf.\ our discussion of
Fig.~\ref{fig,hi_def}c), the two other scenarios deserve some further
attention. H{\sc i} consumption by star formation could help establish
observed deficiencies, but a prerequisite is that the consumed gas is
not simply replenished at a similar rate, for example through the
cooling out of hot, coronal material on to the disc. Our model results
for the X-ray bright groups (Table~\ref{tab,corona}) suggest that such
a strangulation scenario could at least be greatly facilitated by the
removal of coronal gas due to ram pressure, but a more detailed
exploration of model parameter space would be required to assess the
general validity of this conclusion.

In this context, it is instructive to compare the results from our
simple analytical model to the simulation results of
\citet{kawa08}. These authors investigate the efficiency of ram
pressure stripping and strangulation for a galaxy in a small galaxy
group on the basis of a cosmological smoothed particle hydrodynamics
(SPH) simulation. They do not specifically consider a `compact' group,
but this has the advantage that their target galaxy is not subject to
noticeable gas loss from tidal interactions (D.~Kawata, priv.\ comm.),
so in this sense a comparison to our model calculations is
justified. Their target galaxy has initial properties broadly similar
to those of our galaxy model in terms of stellar, gaseous, and total
mass. It is followed for one passage through the group as it enters a
group of virial mass $M \approx 8\times 10^{12}$~M$_\odot$ from an
initial position roughly corresponding to twice the virial radius. The
galaxy orbit thus shows some similarities to our orbit (ii), but, with
a pericentre at $r\sim 200$~kpc, does not take the galaxy through the
very core of the IGM distribution. \citet{kawa08} find that the
resulting ram pressure induces strangulation to a significant
degree. Star formation and thus H{\sc i} consumption becomes mildly
enhanced during infall, and the hot halo gas is almost completely
removed during the group passage.  The combination of these processes
has effectively consumed the H{\sc i} and quenched star formation by
the time the galaxy re-emerges at the virial radius.

Our model conclusion that ram pressure stripping may strongly affect
any coronal gas in the X-ray bright groups is thus in encouraging
agreement with these simulation results, and it is even possible that
our model underestimates the importance of ram pressure in this
context. This is made more relevant still if ram pressure -- or other
galaxy interactions with the group environment -- indirectly
accelerate strangulation by enhancing the disc star formation rate and
hence the consumption of H{\sc i}. This possibility draws
observational support from the strong star formation activity seen in
the ram-pressure affected group spiral NGC\,2276 \citep{rasm06}, as
well as from recent hydro-simulations \citep{kron08}.  However,
\citet{verd98} noted on the basis of star formation rates derived from
{\em IRAS} far-infrared luminosities that there is no indication of
enhanced star formation among the HCG galaxies compared to the level
seen in isolated galaxies. In fact, current star formation rates in
these groups generally seem too low to explain the missing H{\sc i} by
consumption through star formation. For the spirals in our sample, a
mean star formation rate of $\dot{M_\ast}\approx
1.4$~M$_\odot$~yr$^{-1}$ can be derived for the {\em IRAS}-detected
galaxies. However, many of our galaxies remain undetected in one or
more {\em IRAS} bands and so have only upper limits to $\dot{M_\ast}$
(if including these upper limits in the mean, the result is
$\dot{M_\ast} < 0.9$~M$_\odot$~yr$^{-1}$).  At these rates, the
time-scale for star formation to exhaust an initial H{\sc i} supply of
$6.7\times 10^9$~M$_\odot$ to current levels is at least 5~Gyr, even
if neglecting the return of unprocessed material to the ISM {\em and}
any cosmological accretion of gas by the galaxy.

Thus, while our model calculations and the results of \citet{kawa08}
suggest that hot halo gas can be removed fairly efficiently in several
of our groups, potentially inhibiting or at least suppressing any
replenishment of disc H{\sc i} from a hot halo, strangulation does not
seem to have played an important role in establishing current H{\sc i}
levels within our sample. At the observed star formation rates, the
remaining H{\sc i} can continue to fuel star formation for many Gyr
without drastically reducing the H{\sc i} supply. Of course, star
formation rates could have been much higher in the past, which could
also indirectly have contributed to exhausting the gas supply through
the ejection of gas from the disc by starburst winds. However, unless
$\dot{M_\ast}$ in these galaxies has generally declined to current
levels only fairly recently or has been affected by the group
environment over cosmological time-scales, the implication seems to be
that destruction of neutral hydrogen within the galaxies themselves
cannot explain the H{\sc i}~deficiencies. The presence of
intergalactic H{\sc i} in some of our groups also shows that this
scenario cannot provide an exhaustive explanation. We are therefore
compelled to also consider the alternative, externally--driven removal
and destruction of H{\sc i}.

\subsection{H{\sc i} removal by external forces}

Focussing first on the removal of H{\sc i}, both galaxy--IGM and
galaxy--galaxy interactions could be envisaged to play a role. Our
results demonstrate that although earlier studies indicated a link
between significant H{\sc i}~deficiency in Hickson groups and the
presence of hot intragroup gas \citep{verd01}, there is clearly no
one-to-one correspondence between the two. The lack of a detectable
intragroup medium inferred here for half of the eight most H{\sc i}
deficient compact groups specifically appears to rule out IGM
dynamical interactions as generally dominant for H{\sc i} removal
within these environments.  Ram pressure and viscous stripping could
nevertheless still have played a role for H{\sc i} removal in our
X-ray detected systems. Our modelling results indicate, however, that
the efficiency of these processes is generally insufficient to fully
account for the H{\sc i} missing from the individual group members,
especially if our simplistic treatment of viscous stripping
overestimates the predicted H{\sc i} loss, as hypothesized in
Section~\ref{sec,caveats}.

Overall, the simulations of \citet{kawa08} seem to support these
conclusions. In these, ram pressure affects the amount of cold gas in
the disc to an even lesser degree than in our model, although this
could perhaps be attributed to the fact that our groups are at least
twice as massive, and perhaps also to the maximizing orbital
assumptions adopted in our model. Viscous stripping does not appear to
be important in the simulations, but as noted by \citet{kawa08}, such
processes are not necessarily well treated by SPH schemes, so a direct
comparison to our results may not be meaningful.

The limited impact of galaxy--IGM interactions inferred for our groups
commands an alternative explanation for the H{\sc i} removal. Tidal
stripping constitutes an obvious candidate in these dense,
low-$\sigma$ environments, particularly in light of the fact that the
four X-ray undetected groups within our sample exhibit the lowest
velocity dispersions and so could be expected to represent
environments where tidal interactions should be most
important. However, even if for now ignoring the problem of
subsequently heating the removed H{\sc i}, it is not immediately clear
that tidal interactions would necessarily result in increased H{\sc i}
{\em deficiency}, as they would also affect the stellar component in
the galaxies. If stars and gas are removed in roughly equal
proportion, and if the removed stars become part of any undetected
intracluster light (cf.\ \citealt*{gonz07}) while the H{\sc i} remains
detectable, this could potentially even result in an H{\sc i} excess.
Recalling that H{\sc i}~deficiency is defined here on the basis of the
observed $B$-band galaxy luminosity, increasing $\Delta_{\rm HI}$
through tidal stripping therefore requires either the preferential
removal of cold gas compared to stars (and its subsequent heating), or
that $L_B$ is simultaneously boosted relative to the H{\sc i} mass for
the non-stripped components.

The latter possibility gains support from the observation that
specific star formation rates generally tend to be higher in galaxies
with close neighbours \citep{li08b}, with enhanced nuclear star
formation plausibly arising as a consequence of tidally induced gas
inflow. It is perhaps curious then, as mentioned above, that there is
no evidence for enhanced star formation in Hickson groups compared to
the level in isolated galaxies. This runs contrary to expectations for
interacting galaxies, and would seem to argue against $L_B$ being
significantly boosted in the HCG members relative to their remaining
stellar or H{\sc i} mass. The other possibility mentioned above, that
H{\sc i} is more easily tidally stripped than the stellar component,
is perhaps more promising. In many of our groups, such as HCG\,100
(cf.\ Section~\ref{sec,H100}), significant amounts of intergalactic
H{\sc i} is detected with the GBT, whereas the stellar components are
not noticeably affected. A possible explanation is that a relatively
larger fraction of H{\sc i} compared to stars initially resided at
large galactocentric radii, where tidal removal would be most
efficient.  The fact that the H{\sc i} disc in typical spirals and
late-type dwarfs is often more extended than its optical counterpart,
with the radial distribution of H{\sc i} declining less steeply with
radius than that of the $B$-band light \citep{broe97,swat02}, seems to
support such an explanation.

Tidal stripping clearly {\em is} taking place in some of our groups,
notably in the X-ray undetected groups HCG\,100 and HCG\,44. In the
latter, both optical (Fig.~\ref{fig,mosaic}) and H{\sc i} data
(Borthakur et al., in prep.) show strong evidence for the SBc galaxy
HCG\,44d being tidally stripped.  This may indicate that tidal
stripping is the primary mechanism by which H{\sc i} is removed from
the galaxies in these two groups. Although it is tempting to extend
this conclusion to all the X-ray undetected groups in our sample, and
perhaps even beyond, the inconspicuous star formation rates in Hickson
groups in general, and the tentative lack of enhanced nuclear X-ray
activity in the highly H{\sc i}~deficient systems in particular
(Section~\ref{sec,gals}), may not argue in favour of such an
explanation.

\subsection{Destruction of removed H{\sc i}}

As emphasized in the beginning of this Section, H{\sc i}~deficiencies
can only be explained if the hydrogen, once removed from the group
members, is also transformed from its neutral phase. Irrespective of
the processes accomplishing its removal, a mechanism must therefore
also be invoked for ionizing the H{\sc i} during or following its
transfer to intergalactic space. In our X-ray bright groups, a
candidate process is readily available, since any removed H{\sc i} is
expected to evaporate due to heating by the ambient IGM at a rate
$\dot{M} \propto T^{5/2}$, almost independently of realistic values of
the IGM density \citep{spit62}. A detailed comparison of the X-ray and
H{\sc i} properties of these groups should help test this explanation
and will be presented elsewhere (Verdes-Montenegro et al., in
prep.). In the X-ray undetected groups, where any IGM is expected to
be relatively cool (cf.\ Fig.~\ref{fig,hi_def}c) if at all present,
the picture is less clear-cut. Even if these groups do contain an IGM,
the lower predicted IGM temperatures in this subsample imply average
IGM heating time-scales an order of magnitude above those in the X-ray
bright systems, casting doubt on whether this mechanism would be
sufficient.

If gas is predominantly removed by tidal interactions in the X-ray
undetected groups, it is instead conceivable that some of the H{\sc i}
has been heated by tidal shocks, although the presence of
intergalactic H{\sc i} in some of the groups would imply heating
time-scales well in excess of those associated with the H{\sc i}
removal itself. Another possibility is that the column density of any
removed H{\sc i} becomes too low for the gas to be self-shielding
against ionization by cosmic UV radiation. If so, much of the
undetected hydrogen in these groups could potentially be in the form
of a tenuous, photo-ionized intergalactic plasma. A quantitative
investigation of these possibilities is the subject of future work,
but at present a clear picture of the fate of the removed H{\sc i} in
the X-ray undetected groups remains elusive.

\section{Conclusions}\label{sec,conclus}

Based on a sample of eight Hickson compact groups selected for their
high H{\sc i}~deficiencies, we have used {\em Chandra} and {\em
XMM-Newton} data to assess the properties of any hot intragroup medium
(IGM) and constrain the role of galaxy--IGM interactions in removing
H{\sc i} from the galaxies in these groups. The X-ray analysis reveals
a detectable IGM in four of the eight groups. We have tentatively
identified the detected diffuse emission in HCG\,40, a
spiral-dominated group, as associated with an intragroup medium, but
the combination of a low signal-to-noise ratio and an exceptionally
compact galaxy configuration precludes a highly robust conclusion for
this particular system. The remaining three groups are all fairly
X-ray luminous, showing substantial amounts of intergalactic hot gas
with a somewhat disturbed morphology in all cases.

The remarkable X-ray diversity seen across the sample immediately
suggests that the presence of a significant IGM is not a dominant
factor in establishing observed H{\sc i}~deficiencies, despite earlier
results indicating such a connection \citep{verd01}. It is
particularly notable that some of the most H{\sc i}~deficient groups
show no detectable hot IGM, including HCG\,30 which only contains a
few per~cent of the expected H{\sc i} mass for its galaxy content. A
comparison of H{\sc i}~deficiency with either hot IGM mass or
characteristic `mean' ram pressure confirms the lack of a clear
correlation even for the X-ray bright systems, although statistics are
naturally too limited to enable firm conclusions on this basis
alone. The H{\sc i}~deficiency does not seem to depend on IGM
temperature either, suggesting that heat conduction from the IGM does
not play an important role in destroying galactic H{\sc i} (although
once removed, this gas is likely to evaporate on fairly short
time-scales in the X-ray bright groups).

From fitting analytical models to the derived mass profiles of the
X-ray detected groups, we have constructed plausible models of the
gravitational potential and associated radial galaxy orbits for each
of these groups. Combined with the inferred IGM distributions and a
numerical model of a late-type galaxy with properties broadly matching
those of our observed spirals, this has enabled estimates of the
importance of ram pressure stripping and viscous stripping in removing
H{\sc i} from the late-type galaxies in each group. The results
indicate that, even under maximizing assumptions about the galaxy
orbit, ram pressure stripping will remove only small amounts of cold
gas from the group members, peaking at 10--25~per~cent in the X-ray
bright HCG\,97. We find that viscous stripping is generally more
efficient, with the combination of the two processes capable of
removing more than half of the cold ISM in HCG\,97 and potentially
fully accounting for the missing H{\sc i} mass in both HCG\,37 and
HCG\,97. However, the efficiency of viscous stripping is likely
overestimated with our simple analytical approach, and yet these
processes are insufficient in terms of explaining the H{\sc i}
deficiency of the X-ray detected HCG\,15 and HCG\,40.

The model results also indicate that ram pressure can efficiently
remove a large fraction of any hot galactic halo gas that may
otherwise act as a supply of fresh material for star formation in the
disc. However, even if the gas supply to the disc can be completely
cut off, gas consumption at the typical star formation rates in the
groups would proceed far too slowly to explain the observed shortfall
of H{\sc i} by itself. Much higher star formation rates in the recent
past are required for this process to have had any significant
impact. This may suggest that the observed H{\sc i}~deficiencies are
not caused by {\em in situ} destruction of H{\sc i} within the
galaxies themselves. It remains a possibility that even modest star
formation activity could have heated some of the H{\sc i} and lifted
it above the disc midplane where it would be more susceptible to
removal by ram pressure, similar to the situation proposed for
NGC\,2276 \citep{rasm06}. The absence of observational signatures of
this process, e.g., in the form of a hot gas tail extending from any
of the group members, may suggest that such a mechanism is not
generally very important within our sample though.

By the process of elimination, it seems plausible that tidal
interactions have played a key role for H{\sc i} removal in the
groups, particularly in those systems containing no detectable IGM. In
order to explain the H{\sc i} {\em deficiencies}, this scenario would
likely require preferential removal of H{\sc i} over stars, as perhaps
facilitated by the more extended distribution of cold gas relative to
stars in typical late-type galaxies.  While it is perhaps not
surprising that tidal interactions are affecting the gas content of
galaxies in these compact groups, the tidal stripping explanation
still faces some outstanding issues.  Among these is the expectation
that such interactions would generate enhanced star formation or
nuclear activity, but there is no indication that the X-ray faint or
highly H{\sc i}~deficient systems in our sample show evidence for
increased such activity (although the frequency of nuclear activity in
galaxies in compact groups in general may be rather high;
\citealt{mart07}). If interpreting the X-ray bright or highly H{\sc i}
deficient systems within our sample as dynamically more evolved, we
thus find no clear evidence that the frequency or strength of nuclear
X-ray activity in the group members depends on the dynamical status of
the group. It also remains unclear whether tidal interactions
themselves can destroy the H{\sc i} during or following its removal
from galactic discs and so fully explain the H{\sc i}~deficiency in
any of our groups.

In closing, our results suggest that galaxy--IGM interactions can have
played a role for the removal and destruction of H{\sc i} in some of
our groups, but a complete understanding of the origin of the observed
H{\sc i}~deficiencies and the processes causing it is still
lacking. Strangulation or thermal evaporation do not emerge as
important contenders, and typical indirect signatures of tidal
interactions, such as enhanced star formation or nuclear X-ray
activity, are not more pronounced within the more H{\sc i}~deficient
half of our sample.  The latter seems in line with previous results
\citep{verd98} which indicate that star formation rates in Hickson
compact groups are not globally enhanced relative to the field. We
note here that this result could potentially be misleading, however,
perhaps masking an evolutionary trend in which galaxies initially
experience enhanced star formation which is then followed by an
environment--driven suppression. A detailed correlation of H{\sc i}
and X-ray morphology in the groups, coupled with a broad comparison of
individual galaxy properties such as specific star formation rates,
may therefore shed further light on the fate of the missing H{\sc i}
in these compact systems. This is the subject of future work.

\section*{Acknowledgments} 

We thank the referee for useful comments which helped to clarify the
presentation of our results. This work made use of the NASA/IPAC
Extragalactic Database (NED) and the Two Micron All Sky Survey (2MASS)
database.  Support for this work was provided by the National
Aeronautics and Space Administration through Chandra Postdoctoral
Fellowship Award Number PF7-80050 and Chandra Award Number GO5-6127X
and GO6-7128X issued by the Chandra X-ray Observatory Center, which is
operated by the Smithsonian Astrophysical Observatory for and on
behalf of the National Aeronautics and Space Administration under
contract NAS8-03060. LVM is partially supported by DGI Grant AYA
2005-07516-C02-01 and Junta de Andaluc\'{\i}a (Spain).

\bsp 
 
\label{lastpage} 
 
\end{document}